\documentclass{article}
\usepackage[margin=1in]{geometry}
\usepackage{graphicx} 

\usepackage[T1]{fontenc}

\usepackage[affil-it]{authblk}

\usepackage{titlesec}

\usepackage{tikz}
\usetikzlibrary{shapes,arrows}
\usepackage{tkz-euclide}

\usepackage[numbers,sort&compress]{natbib}
\usepackage{amsmath,bm,amsfonts}
\usepackage{xcolor}
\usepackage{MnSymbol}

\usepackage{setspace}

\newcommand{\pdev}[2]{\frac{\partial #1}{\partial #2}}

\newcommand{\grad}[0]{\bm{\nabla}}

\newcommand{\ol}[1]{\overline{#1}}
\newcommand{\ints}[2]{\int_{#1} {#2} \ \mathrm{d}S}
\newcommand{\intv}[1]{\int_{\mathcal{V}} {#1} \ \mathrm{d} V}

\newcommand\Rey{\mbox{\textit{Re}}}  
\newcommand\Ri{\mbox{\textit{Ri}}}  
\newcommand\Pran{\mbox{\textit{Pr}}} 
\newcommand\Frou{\mbox{\textit{Fr}}} 

\newcommand\caseone{$\Rey_{500} \Ri_{005}$}
\newcommand\casetwo{$\Rey_{500} \Ri_{050}$}
\newcommand\casethree{$\Rey_{500} \Ri_{200}$}
\newcommand\casefour{$\Rey_{2000} \Ri_{005}$}
\newcommand\casefive{$\Rey_{2000} \Ri_{050}$}
\newcommand\casesix{$\Rey_{2000} \Ri_{200}$}

\title{Mixing by offshore wind infrastructure: Resolving the density stratified wakes past vertical cylinders}
\author{Charlie J. Lloyd \& Robert M. Dorrell}
\affil{
	School of Architecture, Building and Civil Engineering, Loughborough University, Loughborough LE11 3TU
}

\usepackage[section]{placeins}

\begin{document}

\onehalfspacing

\maketitle

\section*{Abstract}
Offshore wind is rapidly expanding to meet clean and secure energy needs. 
New developments are now increasingly constrained to deeper waters, where the water column is seasonally stratified. 
Here, flows past offshore wind infrastructure will increase water column mixing, although such processes and their extent are poorly understood.
Studies have so far been limited to: field-scale simulations, which make sweeping assumptions regarding flow-structure interactions and fine-scale stratified turbulence; and field observations, which are limited by the sparsity of measurement campaigns and data captured. 
To isolate and quantify the key processes governing water column mixing by infrastructure, we present the first fully structure-resolved direct numerical simulations of two-layer stratified flow past a vertical cylinder.
We identify two wake regimes by systematically varying the flow Reynolds and Richardson numbers: i) A weakly stratified regime, characterised by a narrow but highly energetic wake dominated by horizontal shear, and ii) A strongly stratified wake, characterised by the emergence of a thermocline-spanning recirculation cell attached to the cylinder. Here, strong vertical motions develop from the recirculation cell which are responsible for the formation of large-scale stationary internal waves.
These waves account for up to 10\% of the total energy budget, and provide a new mechanisms for far field energy propagation.
The weakly stratified wake regime is characteristic of existing offshore wind sites where temperature gradients are relatively weak; the newly identified strongly stratified regime describes the dynamics to be expected in new and future deep water offshore wind sites.
This difference between the two regimes explains previously enigmatic field observations regarding wake persistence and detectability.
This work establishes the first mechanistic framework connecting water-column mixing to reversible and irreversible energy exchanges across mean, turbulent, and potential energy reservoirs. The resulting database provides a critical benchmark for validating future models and narrowing the gap between idealized simulations and field-scale flows.

\section{Introduction}
Offshore wind is exponentially expanding across the globe \citep{wang2024remote} to deliver clean, secure, and affordable energy.
Due to spatial constraints in shallow water regions, large-scale developments are now occurring in relatively deep, seasonally stratified, shelf seas, for the first time \citep{dorrell2022anthropogenic}.
Density stratification is a critical control on transport processes in shelf seas, which are disproportionately important for tidal energy dissipation \citep{egbert2000significant}, ocean biological production \citep{wollast1998evaluation}, and the global carbon budget \citep{bauer2013changing}.
This control on transport processes arises through the suppressive nature of buoyancy forces on vertical motion, which can subsequently lead to significant barriers to transport when buoyancy gradients are sufficiently large.

Unlike in shallow water sites, where background tidal/wind-driven processes are able to effectively mix the water column, in deeper waters offshore wind infrastructure introduces a new source of anthropogenic turbulence \citep{carpenter2016potential,dorrell2022anthropogenic,isaksson2025paradigm} which may interfere with the delicate balance between restorative surface heating and water column mixing processes. 
Large-scale build out of offshore wind enhances water column mixing through either atmospheric effects \citep{christiansen2022emergence,daewel2022offshore,zampollo2025does} or directly from tidal flows past infrastructure \citep{rennau2012effect,cazenave2016unstructured,schultze2020increased,hendriks2025impact}. 
Here we focus on the latter mechanism, where wakes shed by tidal flows past infrastructure are a direct source of turbulent motions within the water column.

Field campaigns have attempted to quantify the potential impacts of offshore wind infrastructure on water column mixing
\citep{floeter2017pelagic,schultze2020increased,austin2025enhanced,hendriks2025impact}.
Through water property transects, \citet{floeter2017pelagic} detected a significant weakening of stratification near the centre of two German Bight wind farms (BARD and Global Tech I), which extended up to half the diameter of the ambient tidal excursion. 
Similarly \citet{schultze2020increased} detected a 65\% decrease in the potential energy anomaly in the wake of a 6 m monopile at the German Bight DanTysk wind farm. 
The wake spread to a width approximately 10 times the monopile diameter, but the full extent of the turbulent wake was not captured by this survey, even at a distance over 50 times the monopile diameter \citep{schultze2020increased}.
Field measurements were also obtained by \citet{hendriks2025impact} at the Norther Offshore Wind Farm (Belgian North Sea), in reasonable agreement with the dataset of \citet{schultze2020increased}; Turbulent Kinetic Energy (TKE) measurements were 50\% higher in the wake of a monopile when compared against ambient conditions, and enhanced vertical mixing could be detected in temperature/salinity transects as their vessel passed through the monopile wake. 
Complimentary Large Eddy Simulations (LES) revealed a TKE dissipation rate two orders of magnitude higher in the thermocline than when the monopile was not present; a greater contribution to energy dissipation than the bottom boundary layer \citep{schultze2020increased}. 
Field measurements taken by \citet{austin2025enhanced} verified the dramatic enhancement of energy dissipation in the water column, who measured an order of magnitude increase in TKE dissipation rate downstream of a 4.7 m monopile in the Liverpool Bay Rhyl Flats wind farm, when compared against ambient conditions.
\citet{austin2025enhanced} also estimated the vertical eddy diffusivity, which increased by an order of magnitude in the monopile wake.

These surveys provide evidence that offshore wind infrastructure can lead to enhanced water column mixing; however, variability among these datasets makes direct comparison and reconciliation difficult.
Variability arises due to differences in structure design and size, location, ambient stratification strength, and background turbulence/tidal velocities.
Even within individual studies, issues have arisen regarding the separation between structure-induced mixing processes and natural variability \citep{floeter2017pelagic}, and between repeat surveys, where \citet{schultze2020increased} could not detect a clear signal in temperature transects when revisiting the DanTysk wind farm during a period of stronger stratification. 

The sparsity of field data, both in terms of campaigns and limitations on data collection, make it critical to develop and adopt numerical models capable of predicting impacts of offshore wind infrastructure on the marine environment. 
To date, modelling efforts have mainly focussed on the use of regional-scale oceanographic models.
The studies of \citet{rennau2012effect,cazenave2016unstructured,christiansen2022emergence}, and \citet{christiansen2023large} highlight there are potential regional scale changes to the physical environment induced by flows past offshore wind infrastructure; \citet{christiansen2023large} predict a 10\% change in current velocities and stratification due to the presence of wind farms.
However, these studies highlight a large degree of uncertainty regarding parameterisation of infrastructure wakes, which occur on sub-grid-scales. 
The parameterisation of infrastructure wakes in oceanographic models is based on the closure of \citet{rennau2012effect}, which incorporates physics through
a momentum sink proportional to the drag acting on individual piles, and as source terms in the Reynolds Averaged Navier-Stokes (RANS) turbulence closure: A TKE source term proportional to the power loss to drag, and a corresponding source of TKE dissipation rate.
Subsequently, the closure requires a choice to be made regarding the drag coefficient, and tuning of additional turbulence closure coefficients.
It is unclear how these coefficients should be constrained, due to their dependence on structure design and background flow/stratification conditions.
Original calibration by \citet{rennau2012effect} was carried out by performing structure-resolved simulations of flows past monopile foundations, using a RANS closure. 
However, accurately simulating stratified turbulence is a key challenge in fluid mechanics due to the strong influence of buoyancy on turbulent structures and mixing, leading to highly anisotropic, intermittent, and anti-diffusive flows \citep{caulfield2021layering}. 
Consequently, it is imperative that low-fidelity (e.g. RANS) models are validated against experimental datasets or DNS, due to the sweeping assumptions inherent in eddy-diffusivity closures \citep{caulfield2021layering}.
There is therefore a critical need for robust datasets of stratified-flow interactions with infrastructure, both to benchmark low-cost models and to improve our understanding of the physical mechanisms governing vertical mixing, their dependence on flow and infrastructure properties, and how these insights can be used to constrain regional-scale model closures.

To address this gap, we perform the first fundamental study of the mixing by vertical cylinders in two-layer density stratified flows, as a model for the interactions between offshore wind monopiles and stratified waters.
While there are numerous studies investigating stratified-flow interactions with horizontal cylinders \citep{boyer1989linearly,xu1995turbulent,christin2021fluid}, which induce vertical shear directly in the flow, only \citet{meunier2012stratified,bosco2014three} have performed experiments with vertical cylinders, as part of their studies investigating the effects of cylinder inclination on the onset of two- and three-dimensional instability (i.e. at low Reynolds numbers).
Vertical cylinders introduce primarily horizontal shearing in the water column, and hence, vertical vorticity.
Coherent vertical vortices are known to lead to an efficient route to turbulence in linearly stratified flows, through spontaneous layering \citep{billant2000theoretical,bosco2014three,thorpe2016layers,lucas2019evolution,caulfield2021layering}.
However, the vertical vortices introduced by offshore wind infrastructure are of a similar diameter to that of the thermocline width, both of order 10 m.
This is in contrast to previous experiments, where the cylinder diameter is orders of magnitude smaller than the buoyancy gradient length scale \citep{billant2000theoretical}.
Offshore wind infrastructure in stratified flows may therefore lead to fundamentally different flow mixing processes when compared to previous studies.

We therefore address three key questions: How does the horizontal shear produced by two-layer stratified flow-interactions with vertical cylinders lead to mixing? How are mixing processes affected by the background flow Reynolds number (based on the cylinder diameter) and Richardson number (stratification strength)? And what are the energy pathways that lead to irreversible mixing in such flows? 
These questions are answered using Direct Numerical Simulations at moderate cylinder Reynolds numbers and varying degrees of stratification strength. 
The paper is structured as follows: First, we introduce the methodology in Section \ref{sec:methods}.
In Section \ref{section:results} we present our results, including analysis of the instantaneous flow structure, the time-averaged flow field, the spatial dependence of key transport budgets, and finally assessment of volume-integrated energy budgets. We conclude this study with a discussion in the context of offshore wind marine impacts in Section \ref{section:discussion}. 

\section{Methodology}
\label{sec:methods}
\begin{figure}
    \centering
    \def\scaley{0.5}
\def\spacing{0.05} 
\def\arrowspace{0.1}
\begin{tikzpicture}[x=0.5\textwidth,y=0.5\textwidth*\scaley,scale=0.8]
    \fill[lightgray!60!] (0+\spacing,0.0+\spacing) rectangle (1-\spacing,1-\spacing);
    \draw[draw=black,thick] (0+\spacing,0.0+\spacing) rectangle (1-\spacing,1-\spacing);

    \fill[white]  (0.2,0.5) ellipse (0.04 and 0.04/\scaley);

    \draw[draw=gray,thin,-] (0+\spacing,0.5) -- (1-\spacing,0.5);
    \draw[draw=gray,thin,-] (0.2,0+\spacing) -- (0.2,1-\spacing);

    \draw[thin,-latex] (0.35,0.6) -- (0.45,0.6) node[anchor= north east] {$x$};
    \draw[thin,-latex] (0.35,0.6) -- (0.35,0.6+0.1/\scaley) node[anchor= north east] {$y$};

    \draw[thin,latex-latex] (0.2,0.08) -- (1-\spacing,0.08) node[midway,above] {$40d$};
    \draw[thin,latex-latex] (0.+\spacing,0.08) -- (0.2,0.08) node[midway,above] {$15d$};
    \draw[thin,latex-latex] (0.92,0.5) -- (0.92,1-\spacing) node[midway,left] {$20d$};

    \draw[draw=gray,thin,-] (0.2-0.04,0.5-0.07) -- (0.2-0.04,0.5+0.14);
    \draw[draw=gray,thin,-] (0.2+0.04,0.5-0.07) -- (0.2+0.04,0.5+0.14);
    \draw[thin,latex-latex] (0.2-0.04,0.5+0.1) -- (0.2+0.04,0.5+0.1) node[right,above] {$d\ \ $};

    \draw[draw=black,thick]  (0.2,0.5) ellipse (0.04 and 0.04/\scaley);

    \fill[lightgray!60!] (1+\spacing,0.1) rectangle (2-\spacing,0.9);
    \draw[draw=black,thick] (1+\spacing,0.1) -- (2-\spacing,0.1);
    \draw[draw=black,thick] (1+\spacing,0.9) -- (2-\spacing,0.9);

    \fill[white] (1+0.65,0.1) rectangle (1+0.7,0.9);
    \draw[draw=black,thick] (1+0.65,0.1) rectangle (1+0.7,0.9);

    \draw[thin,-Triangle] (1+0.05,0.2) -- (1+0.15,0.2);
    \draw[thin,-Triangle] (1+0.05,0.2+\arrowspace) -- (1+0.15,0.2+\arrowspace);
    \draw[thin,-Triangle] (1+0.05,0.2+2*\arrowspace) -- (1+0.15,0.2+2*\arrowspace);
    \draw[thin,-Triangle] (1+0.05,0.2+3*\arrowspace) -- (1+0.15,0.2+3*\arrowspace) node[right] {$U_\text{in}$};
    \draw[thin,-Triangle] (1+0.05,0.2+4*\arrowspace) -- (1+0.15,0.2+4*\arrowspace);
    \draw[thin,-Triangle] (1+0.05,0.2+5*\arrowspace) -- (1+0.15,0.2+5*\arrowspace);
    \draw[thin,-Triangle] (1+0.05,0.2+6*\arrowspace) -- (1+0.15,0.2+6*\arrowspace);

    \draw[scale=1, domain=0.1:0.9, smooth, variable=\y, black] plot ({1.4+0.1*tanh((\y-0.5)*10)}, {\y});  
    \node at (1.34,0.2) {$T_0$};
    \node at (1.4,0.8) {$T_0 + \Delta T$};
    \node at (1.48,0.45) {$T(z)$};
    
    \draw[thin,-latex] (1.8,0.4) -- (1.9,0.4) node[anchor= north east] {$x$};
    \draw[thin,-latex] (1.8,0.4) -- (1.8,0.4+0.1/\scaley) node[anchor= north east] {$z$};
        
    \node at (0.05,1.05) {(a)};
    \node at (1.05,1.05) {(b)};
       
\end{tikzpicture}
    \caption{
    	Plan view (a) and side view (b) of the fluid domain and inflow conditions.
    }
    \label{fig:domain}
\end{figure}
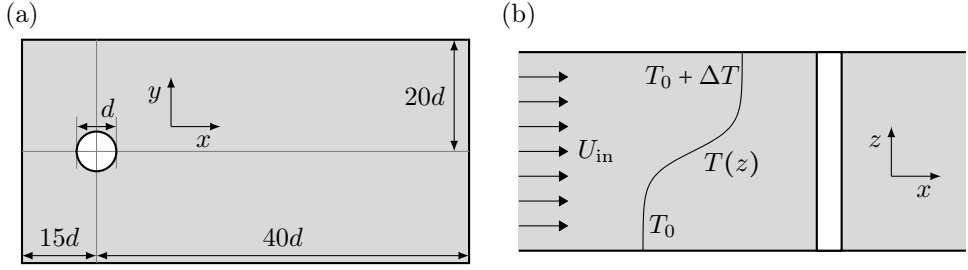

The fluid domain is sketched in Figure \ref{fig:domain},
with $x$ the streamwise coordinate, $y$ the spanwise coordinate, and $z$ the vertical coordinate. 
The cylinder is vertically oriented and placed at the origin $(x,y)=(0,0)$.
The domain extends $15d$ upstream, $40d$ downstream, and $20d$ in the positive and negative spanwise directions. The domain height is set to $10d$ such that the domain coordinates, made dimensionless by the cylinder diameter, are bounded by $x \in [-15,40]$, $y \in [-20,20]$, and $z \in [-5,5]$.

We solve the divergence free, dimensionless and Boussinesq momentum and continuity equations, with a scalar transport equation for a temperature field:
\begin{equation}
\label{eq:mom}
    \pdev{\bm{u}}{t} + \left(\bm{u} \cdot \grad \right) \bm{u}
    =
    - \grad p + \Ri_d (\theta - \theta_\text{in}) \bm{e}_z + \frac{1}{\Rey_d} \nabla^2 \bm{u},
\end{equation}
\begin{equation}
    \grad \cdot \bm{u} = 0,
\end{equation}
and
\begin{equation}
\label{eq:theta_transport}
    \pdev{\theta}{t} + \left(\bm{u} \cdot \grad \right) \theta = \frac{1}{\Pran \Rey_d}\nabla^2 \theta.
\end{equation}
Here, $\bm{u}$ represents the three-dimensional velocity vector, and $\theta = (T - T_\text{ref}) / \Delta T$ is the dimensionless temperature field related to the dimensional temperature $T$, a reference temperature $T_\text{ref}$, and the temperature difference between the upper and lower layers of the flow, $\Delta T$.
$p$ represents the pressure once the approximate hydrostatic pressure has been subtracted, which will be discussed shortly. 
The governing equations have been made dimensionless by the uniform inflow streamwise velocity, $U_\text{in}$, the cylinder diameter $d$, and the temperature difference between the upper and lower layer, $\Delta T$, leading to three dimensionless parameters governing flow dynamics: The cylinder Reynolds number $\Rey_d$, cylinder Richardson number $\Ri_d$, and the Prandtl number $\Pran$:
\begin{equation}
    \Rey_d = \frac{U_\text{in} d}{\nu}, \ \ \Ri_d = \frac{\alpha \Delta T g d}{U_\text{in}^2}, \ \ \text{and} \  \Pran = \frac{\nu}{\kappa},
\end{equation}
where $\alpha$ is the thermal expansion coefficient, $g$ represents gravitational acceleration, $\kappa$ is the diffusivity of temperature, and $\nu$ is the kinematic viscosity. 
This work presents six simulations carried out at two Reynolds numbers ($\Rey_d = 500$ and $\Rey_d = 2000$), one Prandtl number ($\Pran = 1$), and three levels of stratification strength ($\Ri_d = 0.05, 0.5, \text{and} \ 2.0$). 
The six cases are termed \caseone, \casetwo, \casethree, \casefour, \casefive, and \casesix, respectively.

At the upstream (negative $x$) boundary we prescribe $\bm{u} = (1,0,0)$ and
   $ \theta = \theta_\text{in} = \tfrac{1}{2} \tanh{\left( z (d/h) \right)}$,
where $h$ represents the thermocline thickness, which here is set equal to the cylinder diameter: $d/h = 1$.
At the downstream (positive $x$) boundary we apply an open boundary condition \citep{dong2014robust} and no-slip conditions at the cylinder surface, $\bm{u} = 0$.
Note that the pressure field  $p$ has been constructed to be in approximate hydrostatic balance by subtracting $\theta_\text{in}$ from the buoyancy forcing in equation \eqref{eq:mom}.
The full pressure field is recovered by $\tilde{p} = p + \Pi(z)$, where the potential $\Pi(z) = \int_z \Ri_d \theta_\text{in} \mathrm{d} z$.
The remaining boundaries are treated with mixed shear-free boundary conditions: $n_i \partial_i u_j  = 0$, and $u_i n_i = 0$, where $n_i$ represents the surface-normal vector.
The temperature field is treated as insulated on all boundaries other than the upstream $x-$normal boundary: $\partial_i \theta n_i = 0$.

We adopt an overlapping Schwarz-Spectral-Element-Method (SSEM) to discretise the governing equations, implemented in \citet{nek5000}.
The SSEM framework is based on the principles of the overlapping Schwarz method for solving PDEs on overlapping domains \citep{mittal2019nonconforming}. 
In this way a grid of significantly reduced resolution can be adopted in the freestream of the flow, while a high-quality and high-resolution hexahedral mesh can be adopted in the cylinder wake.
The domain of Figure \ref{fig:domain} (a) and (b) is therefore split into two regions: an inner domain and an outer domain, each of which are decomposed into spectral elements on which the governing equations are discretised using a Galerkin method.
The inner domain is bounded by $x \in [-4.39,40]$, $y\in[-5.01,5.01]$, and $z\in[-5,5]$, and discretised using $E_{xy} \times E_z = 3784 \times 40$ elements, while the outer domain is bounded by  $x \in [-15,40]$, $y\in[-20,20]$, and $z\in[-5,5]$, using $E_{xy} \times E_z = 1584 \times 20$ elements.
The two sub-domains overlap by approximately one element
 which enables interpolation between the two domains at run-time.
Equations are solved by means of local approximations based on a high-order tensor-product polynomial basis located at Gauss-Lobatto-Legendre (GLL) nodes.
We adopt $8^3$ GLL nodes in each element (7th order polynomials), which is sufficient to fully resolve turbulent scales for the cases with $\Rey_d = 500$.
In the cylinder wake of the $\Rey_d = 500$ flows, the ratio between the GLL point spacing and the local Kolmogorov length scale has a maximum value of 2.2, with only 15 \% of GLL points with a spacing greater than the local Kolmogorov scale in the inner sub-domain, and none in the outer sub-domain.
At the cylinder wall of the $\Rey_d = 500$ flows, the maximum element size is approximately 4.5 wall-units, with a mean element size of approximately 2 wall units.
These equate to maximum and mean wall-normal GLL node spacings of approximately 0.29 and 0.13 wall units, respectively.

We adopt this same grid for the higher Reynolds number flows ($\Rey_d = 2000$), although note that this resolution is too coarse to resolve all turbulent scales.
In these cases, we adopt modal based explicit filtering to account for unresolved dissipation (See \citet{lloyd2022coupled} and references therein for validation of this technique for stratified turbulent flows).
Unresolved turbulent scales account for approximately 10\% of the volume integrated turbulent kinetic energy dissipation.
Therefore the higher Reynolds number flows are primarily used as a qualitative comparison and speculation as to how dynamics scale with Reynolds number. 

The unsteady governing equations \eqref{eq:mom} to \eqref{eq:theta_transport} are solved in the velocity-pressure form
using the $\mathbb{P}_n\text{-}\mathbb{P}_n$ scheme, where all variables are represented by the same polynomial order \citep{tomboulides1997numerical}.
Equations are solved simultaneously on the two overlapping sub-domains using semi-implicit BDF3/EXT3 timestepping \citep{mittal2021multirate}, with a dimensionless timestep of 1e-3.
Non-linear terms are de-aliased using the $3/2$ rule \citep{orszag1979spectral,canuto2012spectral}.
Boundary data are exchanged between the two overlapping sub-domains by spectral interpolation and 2nd order temporal extrapolation.
At the end of each timestep, a Schwarz iterations is carried out to ensure consistency in the solution in the overlapping region of the two sub-domains.

Cases are initialised with $\bm{u} = 0$ and $\theta = \theta_\text{in}$, and time-stepped until the flow is statistically steady (approximately 500 time units).
Subsequent data collection occurs over 1000 time units to ensure convergence of turbulent kinetic energy and buoyancy variance budgets.

\section{Results}
\label{section:results}
\begin{figure}
\centering
\includegraphics[width=\textwidth]{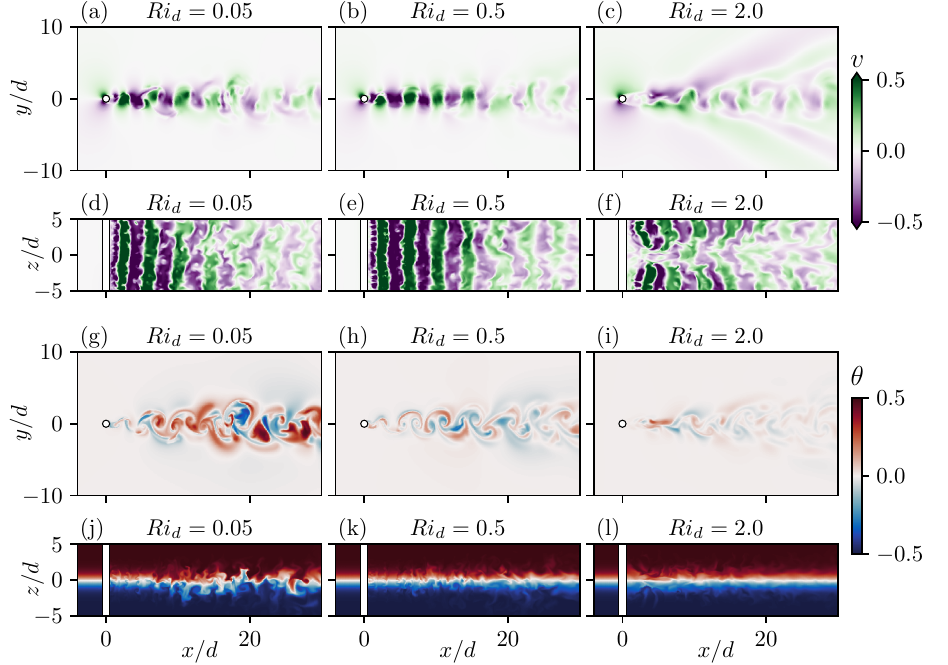}
\caption{
	Visualisation of instantaneous spanwise velocity (a to f) and temperature (g to l). $\Rey_d = 500$ for all panels with: $\Ri_d = 0.05$ in panels (a, d, g, j); $\Ri_d = 0.5$ in panels (b, e, h, k); and $\Ri_d = 2.0$ in panels (c, f, i, l).
	Panels (a) to (c) and (g) to (i) show a $z-$normal slice at $z=0$, and panels (d) to (f) and (j) to (l) show a $y-$normal slice at $y=0$.
}
\label{fig:inst}
\end{figure}

We begin analysis by assessing instantaneous snapshots of the low Reynolds number simulations (note that visualisations of high Reynolds number data can be found in supplementary material).
Visualisations of spanwise velocity and temperature are shown on $y-$ and $z-$normal slices at the origin of the domain, in Figure \ref{fig:inst}.
The cylinder wake is clearly distinct from the uniform background flow.
The Karman vortex (KV) street is most clear in the spanwise velocity signal, which decays in strength for $x \gtrsim 10$.
Qualitatively, the $\Ri_d = 0.05$ and $\Ri_d = 0.5$ flows are very similar, although mixing of the temperature field is clearly suppressed by the higher $\Ri_d$; the large scale vertical mixing structures present in the temperature field for $\Ri_d = 0.05$ are reduced as $\Ri_d$ increases.
Note that for low stratification, mixing of the temperature field appears to grow in strength downstream, while the higher $\Ri_d$ cases suppress vertical motion as the wake evolves.
The temperature perturbations are at their maximum at a streamwise location closer to the cylinder as $\Ri_d$ increases.

The strongest stratification case ($\Ri_d = 2.0$) shows dramatically different behaviour when compared to the other cases. 
While the KV is still present, it is nearly entirely suppressed over the thermocline, and it appears detached from the cylinder surface.
Aside from strong initial fluctuations, the temperature variations for $\Ri_d = 2.0$ are quickly suppressed beyond $x \approx 20$.
A novel and interesting feature of the $\Ri_d = 2.0$ simulation is the large but relatively weak coherent wave-like structures that propagate downstream and away from the energetic KV, most clear in the spanwise velocity signal. 
As we shall show, these are statistically steady coherent structures (stationary internal waves) that appears due to the presence of strong statistically steady vertical motions in the lee of the stratified cylinder wake.

The instantaneous data are qualitatively similar for the high $\Rey_d$ cases, (see supplementary material)
 although the large-scale stationary internal waves appear for $\Ri_d = 0.5$, indicating that the transition between flow regimes is controlled by both $\Rey_d$ and $\Ri_d$. 
The higher Reynolds number cases are more susceptible to the suppression of vertical motion by buoyancy forces due to the earlier emergence and higher importance of three-dimensional flow structures.

\begin{figure}
\centering
\includegraphics[width=\textwidth]{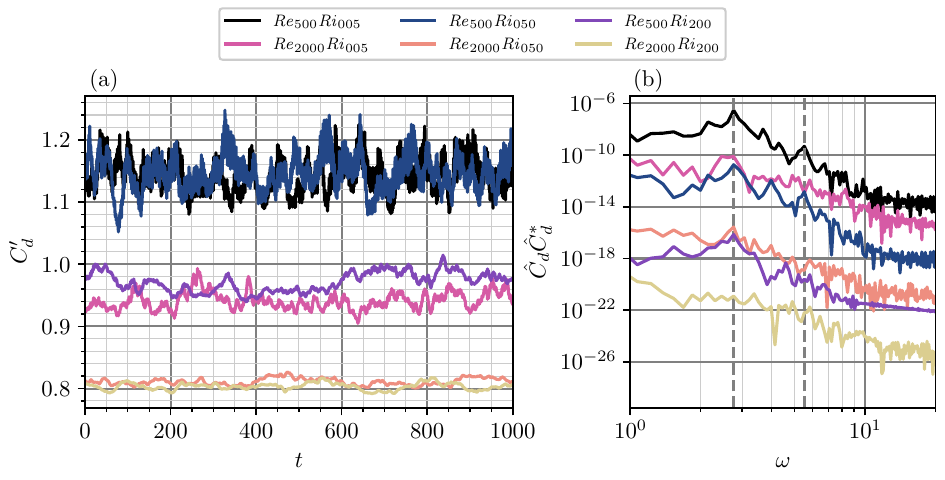}
\caption{
	Drag coefficient time series for all simulations over the full data acquisition time (a).
	Panel (b) shows the power spectral density associated to each drag coefficient time series, with successive cases shifted downward by two orders of magnitude, for clarity.	
	Vertical dashed lines of (b) highlight the dominant vortex shedding frequency ($\omega_\text{peak}$) for case \caseone, and twice this frequency  ($2\omega_\text{peak}$).
}
\label{fig:dragtimeseries}
\end{figure}

The time-dependence of the drag coefficient is presented in Figure \ref{fig:dragtimeseries}, defined by
\begin{equation}
C_d\rq{} = \frac{1}{\frac{1}{2} A_\text{ref} u_\infty^2} \ints{\Omega_w}{\left(p n_x - \frac{1}{\Rey_d} \pdev{u}{x_j} n_j\right)},
\end{equation}
where $n_x$ is the $x-$component of the surface-normal pointing vector (pointing out of the fluid domain), $\Omega_w$ represents the cylinder surface,
and the (dimensionless) normalisation parameters are the reference area $A_\text{ref} = 10$ and the freestream velocity $u_\infty = 1$.
Both the Reynolds number and Richardson number have a strong influence on the drag coefficient.
In agreement with literature data \citep{williamson1996vortex}
the drag coefficient and its fluctuations are reduced as $\Rey_d$ increases from 500 to 2000. 
The reduced fluctuations are reflective of the reduced relative strength of the KV, which is identified by the clear dominant frequency in the power spectral density of the drag coefficient (Figure \ref{fig:dragtimeseries} (b)) at low Richardson numbers, where $\hat{C}_d$ represents the Fourier transform of $C_d\rq{}$ once it\rq{}s mean has been removed, and the superscript $*$ represents the complex conjugate.
At low $\Rey_d$ and $\Ri_d$ there is also evidence of a secondary peak at twice the frequency of the dominant frequency, associated with mode-mode interactions. 
These data also clearly show that as $\Ri_d$ is increased, the drag coefficient and its fluctuations are reduced.
At $\Rey_d = 2000$ and $\Ri_d = 2.0$ there is no convincing peak associated to the KV, and the spectra appears much flatter.
While the KV is still present in the spanwise velocity signal of the $\Rey_d = 2000$ and $\Ri_d = 2.0$ case, it develops further downstream of the cylinder when compared to the other cases, and hence its presence is much weaker in the drag measurements.
\subsection{The mean flow}
Time-averaging is adopted to analyse these psuedo-steady flows, where we adopt an overbar to denote a time-averaged variable, and a prime to denote the fluctuation away from the time-average. 
For example $\bm{u} = \ol{\bm{u}} + \bm{u}\rq{}$.
This section analyses the spatial dependence of the time-averaged fields and key energy transport terms, before we assess surface- and volume-integrated energy budgets.

\begin{figure}
\centering
\includegraphics[width=\textwidth]{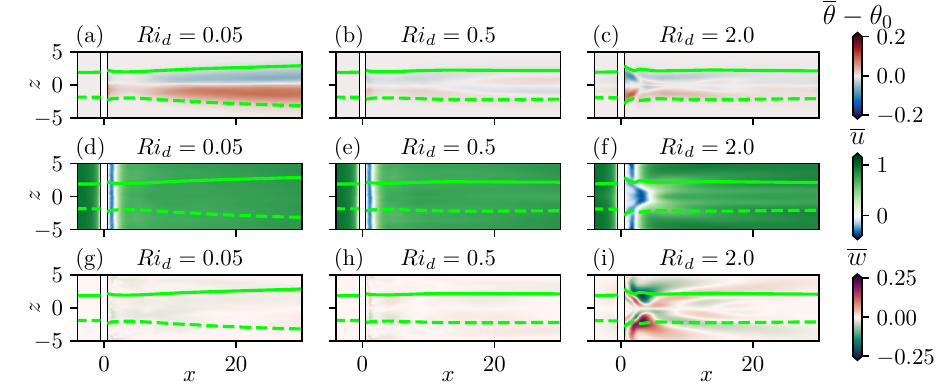}
\caption{
	Time-averaged temperature (a to c), streamwise velocity (d to f) and vertical velocity (g to i) on a $y-$normal slice at $y=0$.
 	$\Rey_d = 500$ for all panels with: $\Ri_d = 0.05$ in panels (a, d, g); $\Ri_d = 0.5$ in panels (b, e, h); and $\Ri_d = 2.0$ in panels (c, f, i).
	Temperature data are presented as a perturbation from the spatially varying background field $\theta_0(x,z)$.
	Lines represent the thermocline bounds, quantified by the contours $\overline{\theta} = \pm 0.475$.
}
\label{fig:meanu_ynormal}
\end{figure}

The vertical dependence of the time-averaged temperature and velocity fields are visualised in Figure \ref{fig:meanu_ynormal}, where data are sampled on a $y-$normal plane at $y=0$.
Here we report the mean temperature field $\ol{\theta}$ with the spatially varying background field $\theta_0(x,z)$ subtracted. 
$\theta_0$ is the temperature field that would arise due to diffusive processes, if there were no obstacle in the flow (obtained by solving the subsequent 2D temperature advection-diffusion equation).
We find that removing this spatially varying reference field enables easier interpretation of temperature perturbations, since any deviations from $\theta_0$ must arise due to the presence of the obstacle. 

\begin{figure}
\centering
\includegraphics[width=\textwidth]{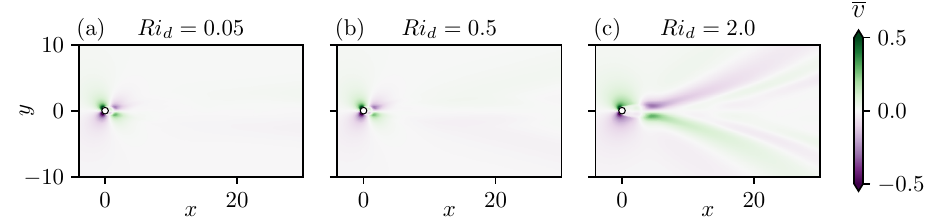}
\caption{
	Time-averaged spanwise velocity on a $z-$normal slice at $z=0$.
 	$\Rey_d = 500$ for all panels with: $\Ri_d = 0.05$ in panel (a); $\Ri_d = 0.5$ in panel (b); and $\Ri_d = 2.0$ in panel (c).
}
\label{fig:meanu_znormal}
\end{figure}

The $\Ri_d = 0.05$ and $\Ri_d = 0.5$ cases appear qualitatively similar for these variables, aside from a stronger broadening of the thermocline for the weakest stratification strength.
The weakest stratification case sees the thermocline broadening over the full domain extent, while the stronger stratification cases appear to suppress thermocline growth.
When stratification is strong ($\Ri_d = 2.0$) there is an intricate recirculation region that develops across the thermocline, leading to strong vertical motion and vertical shear.
This subsequently leads to the stationary waves downstream, observed in the mean vertical velocity, and the temperature deviation.
The stationary waves are also clear in Figure \ref{fig:meanu_znormal}, where the mean spanwise velocity is visualised on a central $z-$normal slice at $z=0$.

It is also interesting to note the sign-changes in the temperature deviation $\ol{\theta} - \theta_0$ (Figure \ref{fig:meanu_ynormal}) as the flow advects downstream for both the $\Ri_d = 0.5$ and $\Ri_d = 2.0$ cases.
This is indicative of reversible energy transfer between the potential and kinetic energy fields, and shows that, while much weaker in magnitude, there is evidence of mean-flow wave-like structures developing in the wake of the $\Ri_d=0.5$ case, similar to the $\Ri_d=2.0$ case but much weaker and further downstream.

\begin{figure}
\centering
\includegraphics[width=\textwidth]{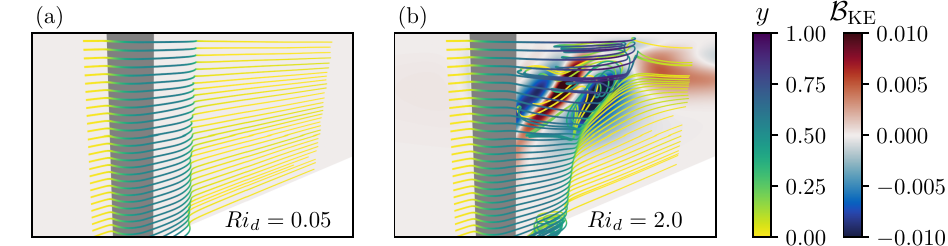}
\caption{
		3D Visualisation of mean flow streamlines in the lower half of the flow, coloured by the spanwise coordinate. 
		$\Rey_d = 500$ for all panels with: $\Ri_d = 0.05$ (a) and $\Ri_d = 2.0$ (b).
		A $y-$normal slice at $y=0$ is added to visualise the mean buoyancy flux, $\mathcal{B}_\text{KE} = - \Ri_d \ol{w} (\ol{\theta} - \theta_\text{in})$.
}
\label{fig:streamlines}
\end{figure}

The recirculating flow region is well visualised by the streamlines of Figure \ref{fig:streamlines}.
While the weak stratification case shows a uniform separation region as the flow passes downstream of the cylinder, the streamlines of the strong stratification case move further laterally in the thermocline, and the recirculation region approximately doubles in size at $z=0$.
The mean kinetic energy buoyancy flux, $\mathcal{B}_\text{KE} = - \Ri_d \ol{w} (\ol{\theta} - \theta_\text{in})$, is also reported here, where a strong negative flux is associated with an upward flow of dense fluid. 
Note that this flux is associated with mean (time-averaged) kinetic energy, as opposed to the turbulent buoyancy flux $\mathcal{B}_\text{TKE} = - \Ri_d \ol{w'\theta'}$ (see Appendix \ref{section:appendix_equations} for details); $\mathcal{B}_\text{KE}$ is non-zero in the strongly-stratification case (\casethree) due to the recirculation region and the stationary waves.
We see that, in the recirculation region, we get alternating patterns of positive/negative mean buoyancy flux associated with downward/upward backflow.
This has a strong local effect on the temperature field (Figure \ref{fig:meanu_ynormal}), although much of the strong changes in temperature are reversed as the flow is convected downstream.  
However, this strong local temperature change, and associated vertical motions, are the source of the large-scale stationary internal waves.

\begin{figure}
\centering
\includegraphics[width=\textwidth]{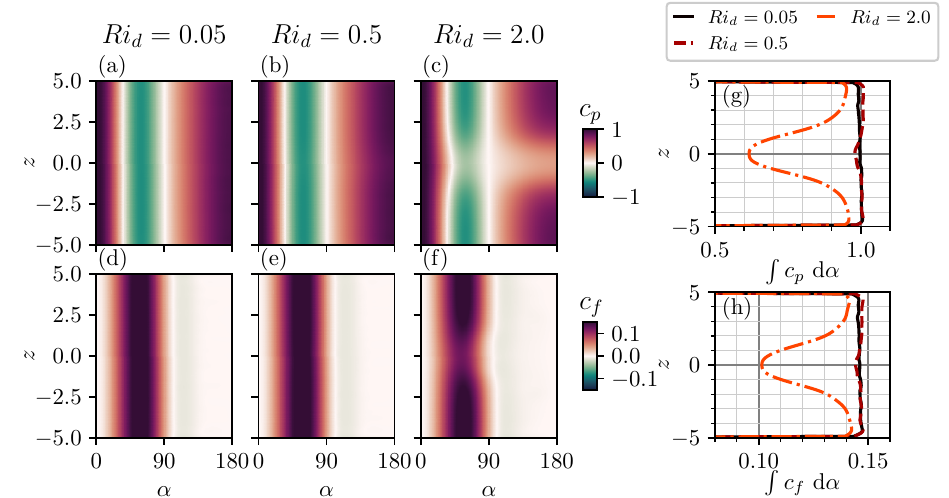}
\caption{
		Spatial dependence of the pressure (a to c) and viscous (d to f) contributions to the drag coefficient as a function of vertical coordinate ($z$) and angular position on the cylinder surface ($\alpha = \arctan{y/x}$).
		$\Rey_d = 500$ for all panels with: $\Ri_d = 0.05$ (a and d); $\Ri_d = 0.5$ (b and e); and $\Ri_d = 2.0$ (c and f).
		Panels (g) and (h) show the pressure and viscous drag coefficients integrated over $\alpha$.
	}
	\label{fig:drag_dist}
\end{figure}

The spatial distribution of the local pressure and viscous drag coefficients over cylinder surface are presented in Figure \ref{fig:drag_dist}, for the three cases with $\Rey_d = 500$.
Here, the local drag coefficient is defined as 
\begin{equation}
	\label{eq:local_drag}
	c_d = c_p + c_\nu = \frac{F_p}{\frac{1}{2} A_\text{ref} u_\infty^2} + \frac{F_\nu}{\frac{1}{2} A_\text{ref} u_\infty^2},
\end{equation}
where $F_p$ and $F_\nu$ are, respectively, the streamwise pressure and viscous forces acting on the cylinder:
\begin{equation}
	F_p = (\ol{p} - p_\infty)n_x, \ \ 
	F_\nu = - \frac{1}{\Rey_d} \pdev{\ol{u}}{x_j} n_j,
\end{equation}
where $p_\infty$ is the reference pressure, equal to the time- and spatially-averaged pressure on the inflow boundary, and $n_j$ is the surface-normal vector pointing out of the fluid volume.
The spatial distributions of the local drag coefficients are presented as a function of the vertical coordinate $z$ and the angular position $\alpha = \arctan y/x$.
At this low Reynolds number we see that viscous contributions to drag are important, and are approximately 15\% of the pressure contributions.
Pressure drag is largest at the upstream and downstream edges of the cylinder surface, while viscous drag peaks at $\alpha \approx 60$ degrees. 
Strong stratification affects both contributions in similar ways when integrated over the angular position shown in Figure \ref{fig:drag_dist} (g,h). 
However, the local effects differ: The upstream side of the cylinder is unaffected by stratification, but the pressure force is significantly reduced across the thermocline. 
In addition, the viscous contributions over the thermocline are reduced, and the region of $c_\nu = 0$ occurs at an earlier angular position, indicative of earlier flow separation and therefore a wider and longer recirculation region (consistent with Figure \ref{fig:meanu_znormal}). 
It is also clear from Figure \ref{fig:drag_dist} that the effects of stratification are felt in the constant density regions, leading to a reduced drag over all $z$. 
This effect would be expected to reduce if the vertical domain height was increased. 
We hypothesise that the local drag reduction across the thermocline is due to the suppression of the KV, and subsequent suppression of spanwise momentum transport, by the large-scale recirculation region that is generated through the strong buoyancy forces across the thermocline.

\begin{figure}
\centering
\includegraphics[width=\textwidth]{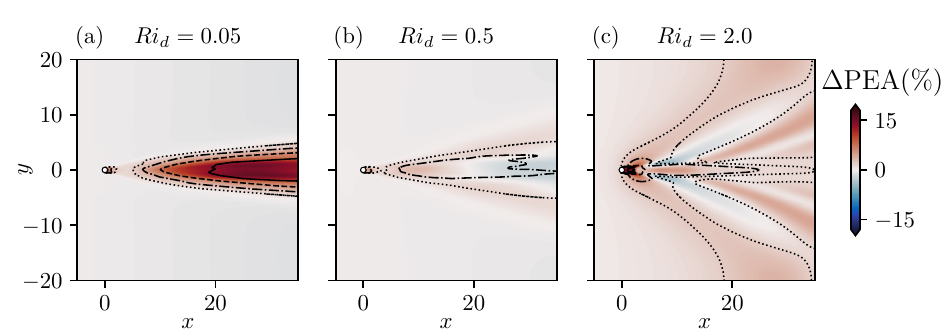}
\caption{
		Change in the Potential Energy Anomaly, defined as $\int_{-2}^2 (E_p - E_{p0}) \  \mathrm{d} z / \left|\int_{-2}^2 E_{p0} \  \mathrm{d} z \right|$.
		$E_{p0}$ represents the spatially varying PE based upon the background temperature field $\theta_0(x,z)$.
		Contours represent the relative changes in thermocline width, defined as the vertical distance between the contours $\overline{\theta} = \pm 0.475$, 
		relative to respective spatially varying background values. 
		Linestyles represent a thermocline width change of 4\% (dotted), 8\% (dash-dotted), 16\% (dashed), and 32\% (solid).
		$\Rey_d = 500$ for all panels with: $\Ri_d = 0.05$ (a); $\Ri_d = 0.5$ (b); and $\Ri_d = 2.0$ (c).
}
\label{fig:pyc_thickness}
\end{figure}

The effect of the cylinder on the potential energy field is expressed as a change in the potential energy anomaly (PEA) in Figure \ref{fig:pyc_thickness}.
Here the PEA is defined as the depth-integrated mean potential energy ($E_p = - \Ri_d \ol{\theta} z$) with the spatially varying background potential energy $E_{p0}$ subtracted, where $E_{p0} = - \Ri_d \theta_0 (x,z) z$:
\begin{equation}
	\Delta \text{PEA} = \frac{\int_{-2}^2 (E_p - E_{p0}) \  \mathrm{d} z }{ \left|\int_{-2}^2 E_{p0} \  \mathrm{d} z \right|}.
\end{equation}
Note that we integrate only over the region where the background temperature field has some appreciable vertical dependence: $-2\leq z \leq 2$. 
This choice is taken since the cylinder cannot affect the fluid $E_p$ in the regions of constant temperature (i.e where the fluid is already mixed), unless the thermocline width has grown sufficiently. 
Contours of the relative thermocline thickness have also been added to Figure \ref{fig:pyc_thickness}: $(\delta^{95} - \delta^{95}_\text{ref}) / \delta^{95}_\text{ref}$, where $\delta^{95}$ represents the vertical distance between the contours $\ol{\theta} = \pm 0.475$, and $\delta^{95}_\text{ref}$ represents the distance between the contours $\theta_0 = \pm 0.475$.
In other words, the thermocline thickness is quantified by the vertical bounds over which 95\% of the total temperature difference is captured.
For reference, $h_\text{ref}$ varies between 3.67 and 4.07 when $\Rey_d = 500$.

With $\Ri_d = 0.05$ we see that both $\Delta \text{PEA}$ and the relative thermocline thickness grow as the turbulent wake is advected downstream, reaching maximum values of $\Delta \text{PEA} \approx 20\%$ and an approximate relative thermocline width of 50\%, although the domain bounds are not large enough to fully capture these true maxima. 
At the strongest level of stratification ($\Ri_d = 2.0$), we see the stationary waves imprinted on the thermocline thickness and PEA, which are at their 
maximum directly behind the cylinder, although these maxima are quickly reversed downstream of the recirculation region. 
When stratification is strong, the presence of the obstacle changes the PEA in the far wake ($x \gtrsim 10$) by approximately $\pm$5-10\%, locally much lower than when stratification is weaker, although this difference is not restricted to just the narrow energetic wake, and is instead more widespread due to the stationary waves. 
A key insight from Figure \ref{fig:pyc_thickness} is that when stratification is strong ($\Ri_d = 2.0$) the thermocline swells and contracts as the stationary wave propagates downstream. 
This is characteristic of solitary `mode 2' internal gravity waves, which 
are thought to be highly efficient at transporting scalars \citep{brandt2014laboratory,carr2015experiments}.

The PEA and $\delta^{95}$ distributions of Case \casetwo\ (Figure \ref{fig:pyc_thickness} (b)) have similarities to both those of \caseone\ (Figure \ref{fig:pyc_thickness} (a)) and \casetwo\ (Figure \ref{fig:pyc_thickness} (c)).
The near cylinder behaviour of \casetwo\ shows a gradual increase in $\Delta \text{PEA}$ and $\delta^{95}$ as the wake propagates downstream, similar to \caseone\ but with a much weaker magnitude.
$\Delta \text{PEA}$ peaks at 10-15 cylinder diameters downstream, before it reduces and becomes negative. 
This is indicative of the prominent internal waves of \casethree, but is of considerably weaker magnitude and hardly detectable in the mean velocity fields (Figures \ref{fig:meanu_ynormal} and \ref{fig:meanu_znormal}).

\begin{figure}
\centering
\includegraphics[width=\textwidth]{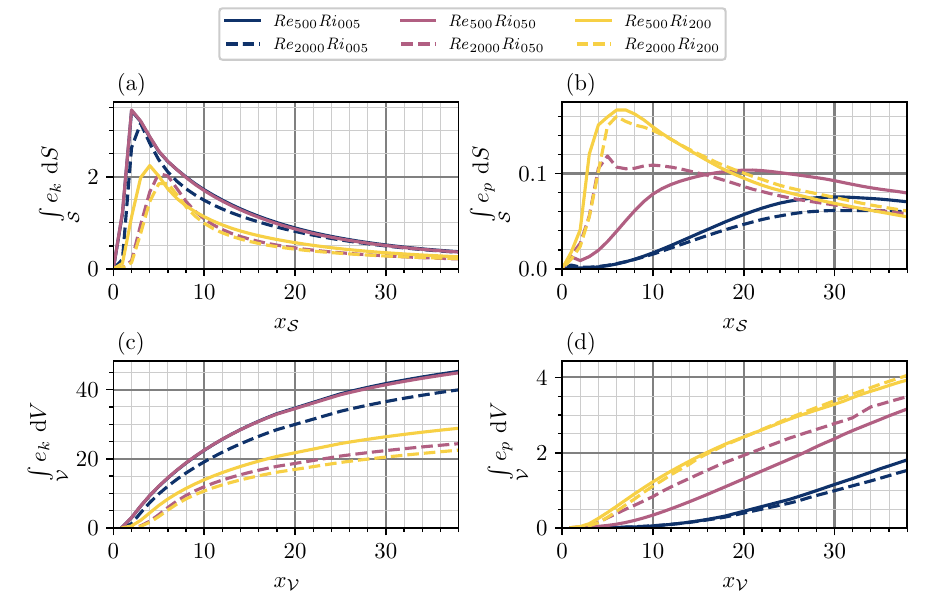}
	\caption{
	Spatial integrals of turbulent kinetic (panels a and c) and potential (panels (b and d) energies, as a function of streamwise position.
	Panels (a) and (b) show integrals over $x-$normal slices at position $x_\mathcal{S}$.
	Panels (c) and (d) show volume integrals over the full domain with a streamwise position less than or equal to $x_\mathcal{V}$.
	}
	\label{fig:energy_integrals}
\end{figure}

The strength of turbulent fluctuations in the wake is visualised in Figure \ref{fig:energy_integrals}, where we report spatial integrals of turbulent kinetic and turbulent potential energies (TKE and TPE), which we define as 
$e_k = \tfrac{1}{2} \ol{u_i\rq{}u_i\rq{}}$ and $e_p = \tfrac{1}{2} \tfrac{\ol{b'b'}}{N^2}$, respectively.
Panels (a) and (b) show integrals of TKE and TPE, respectively, over $x-$normal slices through the full domain as a function of the slice's streamwise position $x_\mathcal{S}$.
The streamwise slice-integrated TKE peaks close to the cylinder surface, where the KV is most dominant. 
This occurs slightly further downstream and with a weaker magnitude for the cases where, across the thermocline, the large-scale recirculation region develops and the KV is suppressed: \casethree, \casefive, and \casesix.
Note that data for \caseone\ and \casetwo\ nearly collapse for the TKE integrals. 
Slice-integrated TPE shows much stronger variation between the six cases, where the peaks in TPE increase with increasing $\Ri_d$.
The cases with $\Ri_d = 0.05$ reach a local maximum at $x_\mathcal{S} \approx 30-35$, and do not decay appreciably before the wake leaves the numerical domain. 
In contrast, while qualitatively similar, case \casetwo\ peaks much earlier at $x_\mathcal{S} \approx 20$, before decaying as the wake propagates further.
A key feature of these flows is that TPE does not peak in the same region as TKE when stratification is weak (\caseone, \casetwo, and \casefour).
The strong-stratification regime of cases \casethree, \casefive, and \casesix, characterised by the vertical recirculation regions and dominant stationary waves, is clearly identified by the rapid increase in slice-integrated TPE at $x_\mathcal{S} \lesssim 4$, reaching their maximum values at $x_\mathcal{S} \approx 6$.
This peak than quickly decays as the wake propagates further, although there is still a reasonable level of TPE downstream. 
Interestingly, the slice-integrated TPE values are similar for all six cases for $x_\mathcal{S} \gtrsim 30$, although differences are likely to emerge if the streamwise domain extent were larger.

Panels (c) and (d) of Figure \ref{fig:energy_integrals} show the TKE and TPE integrated over the domain volume which is bounded by $[x_\text{min},x_\mathcal{V}]$ in the streamwise direction. 
The plateau of these various curves gives an indication of the required domain size for fluctuating processes in the cylinder wake to be fully captured.
The rate of change of these volume integrals with respect to $x_\mathcal{V}$ of course corresponds to the slice-integrals of panels (a) and (b).
The volume integrated TKE (Figure \ref{fig:energy_integrals} (c)) rapidly increases in the near-cylinder wake, where the KV is present, before beginning to flatten out further downstream.
While the domain extent is not long enough to capture the full turbulent wake (which would be identified by an invariant TKE volume integral with respect to $x_\mathcal{V}$), it does capture the most energetic regions.
The total TKE decreases with both increasing $\Rey_d$ and increasing $\Ri_d$, the latter partly due to the suppression of the KV.
This is also consistent with the reduction in drag coefficient with both $\Rey_d$ and $\Ri_d$ (see Figure \ref{fig:dragtimeseries}).
The volume-integrated TPE (Figure \ref{fig:energy_integrals} (d)) increases with increasing $\Ri_d$, and shows only a small degree of convergence as $x_\mathcal{V}$ increases, and only for the strongest stratification cases.
From slice- and volume-integrated data it is clear that significant TPE changes are present further downstream in the wake than TKE, relative to respective maxima.
This is due to the substantially stronger near-cylinder peak of TKE when compared to TPE.
As a result, slice-integrated TKE decreases by approximately a factor of six over the domain extent while TPE only decreases by approximately a factor of two.
The subsequent effect of this is that the rate of change of TKE, with respect to downstream distance changes much more significantly, and exhibits some degree of convergence, while TPE is still increasing substantially over the chosen domain length. 

\begin{figure}
\centering
\includegraphics[width=\textwidth]{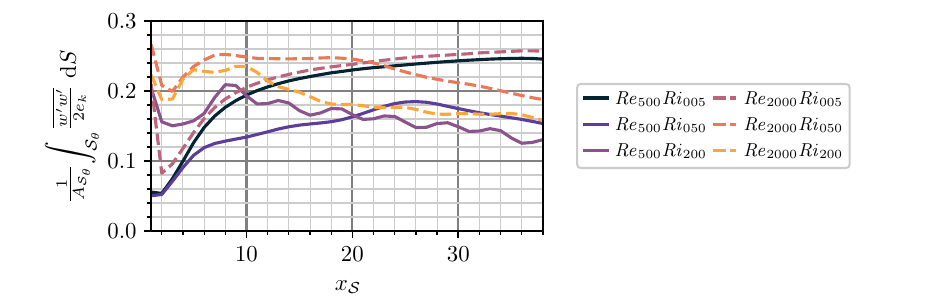}
	\caption{
		Anisotropy in the stratified cylinder wake, visualised by the ratio between the vertical component of TKE against the sum of all three components.
		This anisotropy measure is computed and spatially averaged on $x-$ normal slices at the streamwise locations $x_\mathcal{S}$, within the thermocline bounds $-0.475 \leq \overline{\theta} \leq 0.475$.
		This surface is denoted $\mathcal{S}_\theta$ with an area $A_{\mathcal{S}_\theta}$.
	}
	\label{fig:anisotropy}
\end{figure}

A key result here is that, for weaker levels of stratification, fluctuations in potential energy peak much further from the cylinder than TKE, and therefore the volume-integrated TPE is far from convergence when compared to TKE. 
We hypothesise that this is associated with the transition between horizontally oriented turbulent fluctuations to isotropic turbulence further downstream, which is more readily able to mix the buoyancy field.
To support this argument we inspect the ratio $\ol{w'w'}/2 e_k$ as a measure of flow anisotropy.
This ratio is reported in Figure \ref{fig:anisotropy}, averaged over $x-$normal slices as a function of streamwise position, $x_\mathcal{S}$, within the thermocline bounds $-0.475 \leq \overline{\theta} \leq 0.475$.
The behaviour of flow anisotropy is different for both flow regimes.
In the weak stratification regime (\caseone, \casetwo, and \casefour) we see that vertical fluctuations contribute little to TKE, but they slowly grow, as the wake advects downstream. 
This approximately correlates to the growth of TPE (Figure \ref{fig:energy_integrals} (b)), and confirms that as anisotropy in the wake decreases, temperature fluctuations increase in strength.
Note that, in this numerical domain, isotropic turbulence is never reached, with vertical fluctuations contributing at most 25\% of the total TKE (with 33\% representing an equal contribution to horizontal components).
In contrast, when stratification is strong (\casethree, \casefive, and \casesix) the TKE has a much stronger contribution from vertical fluctuations in the near-cylinder wake. 
This is due to the strong shearing generated in and near the recirculation region.
This then decays due to suppression from buoyancy forces as the wake advects downstream, correlating to the decrease in TPE (Figure \ref{fig:energy_integrals} (b)).
These results demonstrate the close connection between the ability for the flow to suppress vertical fluctuations in the wake, and fluctuations in the buoyancy field.

\subsection{Local energy budgets}
\begin{figure}
\centering
\includegraphics[width=\textwidth]{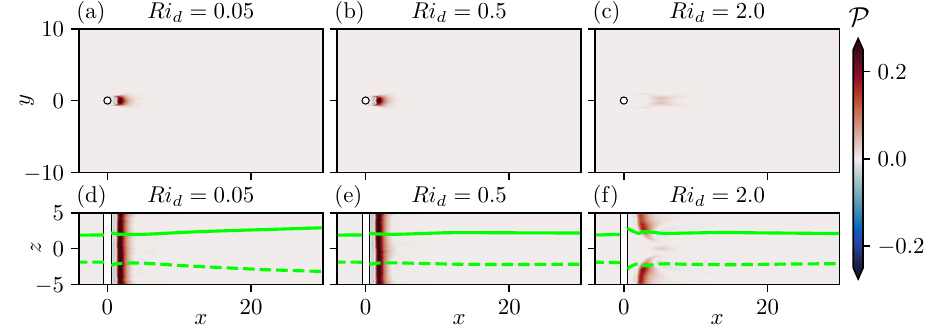}
\caption{
	Visualisation of turbulent kinetic energy production on a $z-$normal slice at $z=0$ (panels a to c) and a $y-$normal slice at $y=0$ (panels d to f).
 	$\Rey_d = 500$ for all panels with: $\Ri_d = 0.05$ in panels (a) and (d); $\Ri_d = 0.5$ in panels (b) and (e); and $\Ri_d = 2.0$ in panel (c) and (f).
}
\label{fig:TKE_prod}
\end{figure}

\begin{figure}
\centering
\includegraphics[width=\textwidth]{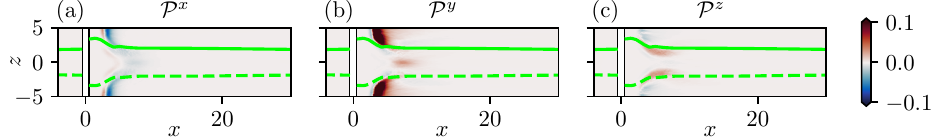}
\caption{
	Components of turbulent kinetic energy production, split by the direction of shear, for $\Rey_d = 2000$ and $\Ri_d = 2.0$. 
	Panel (a) reports the streamwise shear term $\mathcal{P}^x$,
	panel (b) reports the spanwise shear term $\mathcal{P}^y$,
	and panel (c) reports the vertical shear term $\mathcal{P}^z$.
}
\label{fig:TKE_prod_dist}
\end{figure}

Production of TKE ($\mathcal{P} = - \ol{u_i'u_j'} \partial_i \ol{u}_j$) is shown on spanwise and vertical slices for the three cases \caseone, \casetwo, and \casethree, in Figure \ref{fig:TKE_prod}.
TKE production is localised to the near-cylinder wake, and peaks where the KV is strongest.
For $\Ri_d = 0.05$ and $\Ri_d = 0.5$ this occurs within the first few diameters downstream of the cylinder.
For $\Ri_d = 2.0$, TKE production varies with depth, and is small within the thermocline bounds.
TKE production is primarily associated with spanwise shear due to the $z-$invariant cylinder cross section.
Indeed, for cases \caseone\ and \casetwo\ the spanwise shear dominates over other contributions to $\mathcal{P}$, particularly since there is no mean vertical shear. 
However, there is a transition away from this dominance of spanwise shear when stratification is strong, as we show in Figure \ref{fig:TKE_prod_dist} for \casesix.
Here we have split $\mathcal{P} = -\ol{u_i'u_j'} \partial_i \ol{u}_j$ into three contributions which respectively group the nine terms containing streamwise ($x$), spanwise ($y$), and vertical ($z$) spatial derivatives (e.g. $\mathcal{P}^x = -\ol{u'u_j'} \partial_x \ol{u}_j$).
In the unstratified regions, spanwise shear terms ($\mathcal{P}^y$) dominate over other sources of TKE production.
However, within the thermocline bounds, $\mathcal{P}^y$ reduces significantly and vertical shear terms ($\mathcal{P}^z$) become important. 
For this particular case (\casesix), vertical shear terms contribute more to TKE production than spanwise shear terms, within the thermocline.
This importance of $\mathcal{P}^z$ is consistent across all `strongly stratified' wake cases (\casethree, \casefive, and \casesix) although we note that this effect is most pronounced for \casesix.
These results indicate that there is a transition in how turbulence is produced within the thermocline, as a function of stratification strength.
If stratification is strong enough to generate a large-scale recirculation region, then this can lead to a new mechanism, via vertical shearing, for turbulence production. 

\begin{figure}
\centering
\includegraphics[width=\textwidth]{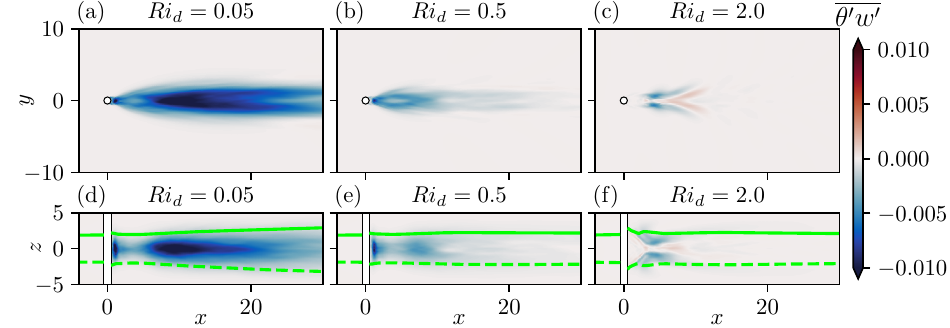}
\caption{
	Visualisation of the vertical temperature flux on a $z-$normal slice at $z=0$ (panels a to c) and a $y-$normal slice at $y=0$ (panels d to f).
 	$\Rey_d = 500$ for all panels with: $\Ri_d = 0.05$ in panels (a) and (d); $\Ri_d = 0.5$ in panels (b) and (e); and $\Ri_d = 2.0$ in panel (c) and (f).
}
\label{fig:buoy_flux}
\end{figure}

The vertical temperature flux ($\overline{\theta'w'}$) is shown in Figure \ref{fig:buoy_flux}.
This is related to the turbulent buoyancy flux by $\mathcal{B}_\text{TKE} = - \Ri_d \overline{\theta'w'}$.
For \caseone, the temperature flux peaks much further downstream ($x \approx 10$) than $\mathcal{P}$, although the peak is less localised.  
There is also a small local peak in $\overline{\theta'w'}$ in the recirculation region near the cylinder surface, although most temperature flux occurs further downstream. 
There are similarities between the temperature flux variations for \caseone\ and \casetwo, although the peak occurs closer to the cylinder when stratification is stronger.
We hypothesise that, for these two cases, where spanwise shear dominates over other sources of TKE production, there requires some transition from spanwise motion to vertical motion as the wake propagates downstream in order for the temperature field to mix. 
Indeed, this is reflected in the anisotropy discussions of Figure \ref{fig:anisotropy}, where vertical fluctuations grow as the wake propagates downstream.
For this reason, the turbulent buoyancy flux peaks much further downstream than TKE production.

In contrast, $\overline{\theta'w'}$ is much more localised when stratification is strong, peaking in the region where $\mathcal{P}$ is large, particularly $\mathcal{P}^z$ (Figures \ref{fig:TKE_prod} and \ref{fig:TKE_prod_dist}).
We hypothesise that this arises due to the strong vertical shearing that produces turbulence for this case, which is more efficiently able to generate vertical motions and therefore buoyancy flux. 
	Note, however, that at this level of stratification, despite the buoyancy flux being much larger, the overall influence on the thermocline thickness (Figure \ref{fig:pyc_thickness}) is weaker due to the strong buoyancy forces. 
However, there are clear regions of negative buoyancy flux downstream of the local peaks for \casethree.
Subsequently, some energy lost to potential energy is transferred back to TKE further downstream.
In other words, there are reversible processes within $\mathcal{B}_\text{TKE}$.

\begin{figure}
\centering
\includegraphics[width=\textwidth]{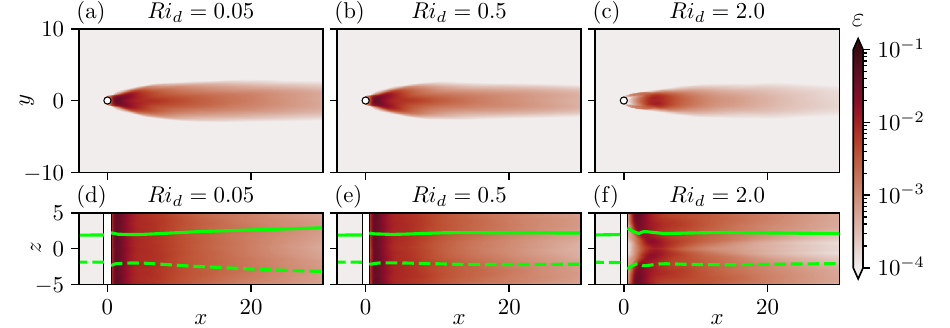}
\caption{
	Visualisation of turbulent kinetic energy dissipation on a $z-$normal slice at $z=0$ (panels a to c) and a $y-$normal slice at $y=0$ (panels d to f).
 	$\Rey_d = 500$ for all panels with: $\Ri_d = 0.05$ in panels (a) and (d); $\Ri_d = 0.5$ in panels (b) and (e); and $\Ri_d = 2.0$ in panel (c) and (f).
}
\label{fig:eps}
\end{figure}

\begin{figure}
\centering
\includegraphics[width=\textwidth]{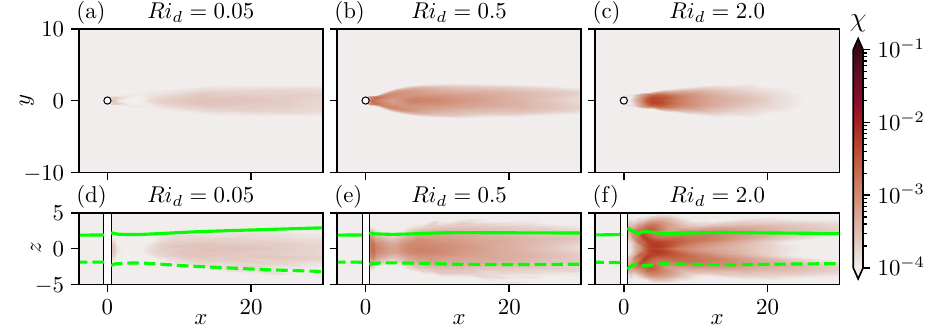}
\caption{
	Visualisation of turbulent potential energy destruction rate on a $z-$normal slice at $z=0$ (panels a to c) and a $y-$normal slice at $y=0$ (panels d to f).
 	$\Rey_d = 500$ for all panels with: $\Ri_d = 0.05$ in panels (a) and (d); $\Ri_d = 0.5$ in panels (b) and (e); and $\Ri_d = 2.0$ in panel (c) and (f).
}
\label{fig:chi}
\end{figure}

Irreversible mixing is discerned through visualisations of $\varepsilon$, the destruction rate of TKE, in Figure \ref{fig:eps}, and $\mathcal{X}$, the destruction rate of scaled buoyancy variance, in Figure \ref{fig:chi}.
These quantities are defined
\begin{equation}
	\varepsilon = -  \frac{1}{Re}\ol{ \pdev{u'_i}{x_j} \pdev{u'_i}{x_j}}, \ \text{and} \  
	\mathcal{X} = - \frac{1}{N^2 \Pran \Rey_d}\overline{\pdev{b'}{x_j}\pdev{b'}{x_j}},
\end{equation}
with the buoyancy defined as $b = \Ri_d \theta$  (See Appendix \ref{section:appendix_equations} for details of respective transport equations).
$\varepsilon$ is primarily focused in the near-cylinder wake with $x \lesssim 5$ (note the log-scaled colourbar), and concentrated in a narrow region downstream.
The strong stratification of \casethree\ weakens the maximum of the dissipation rate and shifts it further downstream, due to the recirculation region. 
Vertical dependence of $\varepsilon$ develops, where it is suppressed in the thermocline as the wake advects downstream.
The magnitude of $\mathcal{X}$ increases with increasing $\Ri_d$, but more interestingly it does not peak in the same regions as the TKE dissipation rate when stratification is weak; for \caseone, $\mathcal{X}$ rises slowly as the wake propagates and doesn't reach an appreciable maximum. For \casetwo, the maximum is reached at $x\approx 10$ before $\mathcal{X}$ decays downstream. 
When stratification is stronger (\casethree), there is a more clear correlation between $\varepsilon$ and $\mathcal{X}$, which both peak near the recirculation region and decay downstream, leading to a far-wake regime ($x \gtrsim 20$) where most TKE and TPE destruction occurs near the bounds of the thermocline. 

Mixing in stratified flows is often parameterised as a function of three dimensionless parameters (excluding the Prandlt number $\Pran$): $\Ri_g$, $\Frou_h$, and $\Rey_b$, which respectively represent the gradient Richardson number, the horizontal Froude number, and the buoyancy Reynolds number \citep{caulfield2021layering}:
\begin{equation}
	\label{eq:dimless}
	\Ri_g = \frac{N^2}{S^2}, \ \ \Frou_h = \frac{\varepsilon}{N u_\text{rms}^2}, \ \text{and} \ \ 
	\Rey_b = \Rey_d \frac{\varepsilon}{N^2}.
\end{equation}
Here, $S = \partial_z \overline{u}$ represents vertical shearing in the streamwise flow.
$\Ri_g$ represents the ratio between buoyancy forces and mean vertical shear, and is a measure of a flows susceptibility to shear instability.
The horizontal (turbulent) Froude number characterizes the effect of stratification on turbulent structures. Large values indicate that buoyancy has little influence while small values ($\Frou_h \ll 1$) indicate that stratification suppresses vertical motions, leading to strongly anisotropic, layer-like “pancake” structures.
The buoyancy Reynolds number represents the ratio of the turbulent kinetic energy dissipation to the damping effects of viscosity and stratification. It can also be interpreted as the square of the ratio between the buoyancy timescale and the Kolmogorov timescale. Large 
  values indicate that turbulence can sustain an energy cascade despite stratification, while small values ($\Rey_b \lesssim 20$)
 imply that stratification and viscosity suppress small-scale turbulent motions, limiting mixing \citep{smyth2000length}.

\begin{figure}
\centering
\includegraphics[width=\textwidth]{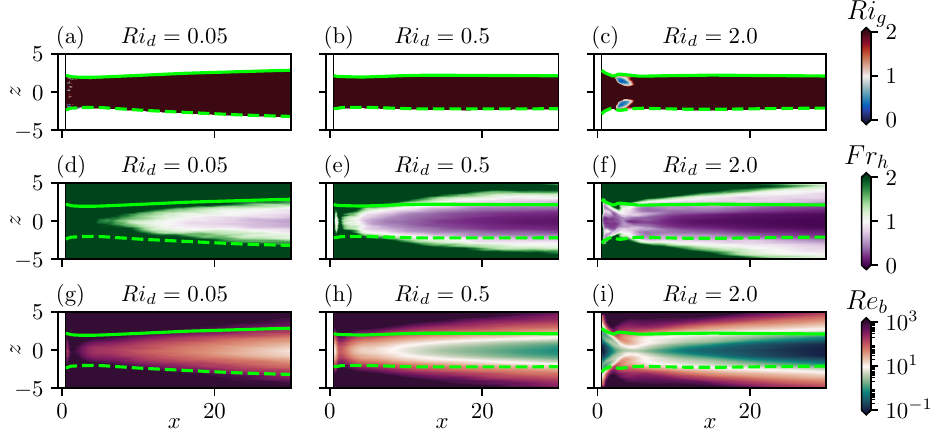}
\caption{
	Key dimensionless parameters for parameterising mixing in stratified flows on a $y-$ normal slice at $y=0$.
	Panels (a to c) show the gradient Richardson number,
	panels (d to f) show the horizontal Froude number, 
	and panels (g to i) show the buoyancy Reynolds number, 
	all defined in Equation \eqref{eq:dimless}.
	Panels (a,d,g) show data for $\Ri_d = 0.05$,
	panels (b,e,h) show data for $\Ri_d = 0.5$,
	and panels (c,f,i) show data for $\Ri_d = 2.0$.
	Lines represent the approximate thermocline width with contours of $\overline{\theta} = \pm 0.475$.
}
\label{fig:dimless}
\end{figure}

These three parameters are visualised on $y-$normal slices at $y=0$ in Figure \ref{fig:dimless}.
First, note that $S^2$ is extremely small over much of the domain, and $N^2$ also approaches small values outside the thermocline.
For this reason, we have limited visualisation of $\Ri_g$ to the thermocline bounds where at least $N^2$ is large. 
Figures \ref{fig:dimless} (a) to (c) clearly demonstrate that $\Ri_g$ is very large in the stratified wakes, due to the lack of vertical mean shear.
There are however two small (vertically symmetric) regions with $\Ri_g < 1$ in the strongly stratified wake of \casethree, which are downstream of the recirculation region and just beneath the thermocline bounds. 
These correlate to the regions where $\mathcal{P}^z$, $\mathcal{B}_\text{TKE}$, and $\mathcal{X}$ are large (Figures \ref{fig:TKE_prod_dist}, \ref{fig:buoy_flux}, and \ref{fig:chi}), confirming that it is indeed a susceptibility to vertical shearing, due to the vertical recirculation zone, that promotes mixing of the temperature field in the strongly stratified wake. 
While $\Ri_g$ is not a useful parameter for characterising the weakly stratified flow, it is useful in characterising the transition from turbulence produced from horizontal shear to vertical shear. 

$\Frou_h$ and $\Rey_b$ are clearly correlated when inspecting Figure \ref{fig:dimless}.
The weakest stratification case (\caseone) shows that $\Rey_b > 10$ and $\Frou_h \gtrsim 1$ for the whole flow, with $\Rey_b > 1000$ in the near-cylinder wake and unstratified regions.
Within the thermocline, both parameters gradually decrease as the wake propagates downstream and turbulence weakens.
\casetwo\ behaves in a similar fashion, although finds that for $x \gtrsim 10$ the horizontal Froude number and the buoyancy Reynolds number are appreciably small to suggest there is a strong influence of buoyancy on turbulent structures, and mixing processes weaken downstream.
This is consistent with the region where $\mathcal{X}$ peaks in the wake before being suppressed downstream (Figure \ref{fig:chi}). 
\casethree\ differs from the other cases, where $\Rey_b$ and $\Frou_h$ are everywhere small in the cylinder stratified wake, aside from the small region at $x \approx 5$, where $\Frou_h \approx 1$ and $\Rey_b \approx 100$. 
This region is unsurprisingly correlated to that of the high $\Ri_g$ region; vertical shearing leads to localised relatively strong turbulence in the near-cylinder wake, which is suppressed as the wake propagates downstream. 

The anisotropy measure of Figure \ref{fig:anisotropy} is correlated to the horizontal turbulent Froude number.
We see that weak stratification has $\Frou_h \gtrsim 1$ across the majority of the wake, and so turbulence anisotropy, initially strong due to the dominance of spanwise motion (low $\ol{w'w'}/2 e_k$), increases as the wake propagates.
$\ol{w'w'}/2 e_k$ plateaus at the same region where $\Frou_h \sim 1$.
Behaviour is similar at moderate stratification (\casetwo), although the $\ol{w'w'}/2 e_k$ reaches its maximum value much earlier before decaying downstream, correlating where $\Frou_h \lesssim 1$.
When stratification is strongest (\casethree) we see $\Frou_h$ is small throughout the whole wake, aside from the region where $\Ri_g < 1$, corresponding to turbulence production.
We therefore see higher isotropy near the cylinder with a maximum value in this region of turbulence production, before vertical motions are suppressed downstream. 

The dimensionless parameters $\Ri_g$, $\Frou_h$, and $\Rey_b$ therefore characterise the two flow regimes: At weak stratification we see $\Ri_g$ is everywhere large due to the negligible vertical shear, and the near-cylinder wake is characterised by a highly turbulent (large $\Rey_b$) flow only weakly affected by buoyancy ($\Frou_h \gtrsim 1$). 
Further downstream this gradually transitions to a buoyancy affected wake once turbulence decays, with lower $\Rey_b$ and $\Frou_h$. 
Mixing of the temperature field occurs in the far-wake, once vertical turbulence motion has developed due to the gradual increase in isotropy, away from the dominant horizontal motion in the near-wake. 
The strong stratification regime is characterised by low $\Rey_b$ and low $\Frou_h$ everywhere in the stratified wake, which leads to the generation of strong vertical shearing associated with vertical secondary motions near the thermocline bounds.
This generates a localised region of low $\Ri_g$ where turbulence in the stratified wake can be produced through vertical shearing, creating a locally energetic (high $\Rey_b$) flow only weakly affected by buoyancy (high $\Frou_h$). 
As the wake propagates, turbulence is quickly suppressed and anisotropy re-develops. 
The dimensionless parameters of Figure \ref{fig:dimless} highlight that the transition from the weakly stratified to the strongly stratified wake is gradual, where, consistent with $\Delta \text{PEA}$ (Figure \ref{fig:pyc_thickness}), elements of both regimes are present for \casetwo.
However, the strongly-stratified wake regime has not yet developed, since the near-cylinder flow is still dominated by horizontal shear.
The following section will assess volume-averaged energy budgets to quantify how these flow regimes affect mixing processes in the water column.

\subsection{Volume-Integrated Energy Budgets}
Of key interest to the offshore wind sector is the fate of energy introduced in the water column.
This energy pathway can be identified through a volume-integrated approach of the transport equations for: Mean Kinetic Energy (KE), Mean Potential Energy (PE), Turbulent Kinetic Energy (TKE), and the Scaled Buoyancy Variance (SBV), which are detailed in Appendix \ref{section:appendix_equations}, \eqref{eq:KE_transport} to \eqref{eq:bb_transport}.
Some degree of volume-averaging is required to properly partition these various budgets, particularly due to the larger instantaneous transport terms inevitable in this advecting flow.
Energy budgets are constructed by integrating over volume $\mathcal{V}(x_\mathcal{V})$, defined as
\begin{equation}
	\mathcal{V} = \int_\mathcal{V} \mathrm{d}V =  \int\displaylimits_{x_\text{min}}^{x_\mathcal{V}} \int\displaylimits_{y_\text{min}}^{y_\text{max}} \int\displaylimits_{z_\text{min}}^{z_\text{max}}  \mathrm{d} z\  \mathrm{d} y\  \mathrm{d} x,
\end{equation}
where we integrate over the full numerical domain up to the streamwise position $x_\mathcal{V}$. 
In this way, we construct volume-integrated budgets as a function of how far the energetic wake was propagated, and we can determine how well turbulent/mixing processes are resolved by their dependence on $x_\mathcal{V}$, which is critical for understanding how far stratified wakes may propagate in the field. 

Integration of the mean kinetic energy (KE) budget \eqref{eq:KE_transport} over $\mathcal{V}$ leads to
\begin{equation}
	\label{eq:partial_KE_vol_int}
	 0 =  
	 	\intv{ - \mathcal{P} - \mathcal{B}_\text{KE} - \mathcal{E} } 
	   +\ints{\Omega}{
		 - \ol{u}_j E_k n_j  - \ol{u}_j \ol{p} n_j + \frac{1}{\Rey_d}  \pdev{E_k}{x_j} n_j -  \ol{u}_i \ol{u_i'u_j'} n_j
		},
\end{equation}
where $\Omega$ represents the bounding surface of $\mathcal{V}$ with the out-ward pointing surface-normal $n_i$, $E_k= \ol{u}_i \ol{u}_i / 2$ represents KE, and $\mathcal{E}$ represents the KE destruction rate (typically small for high Reynolds number flows).
The transport terms can be manipulated to identify the KE source term as the power loss to drag \citep{gao2019energy,bonnavion2022use}.
This is achieved by first splitting the bounding surface into $\Omega = \Omega_i + \Omega_w + \Omega_o$, which respectively represent the in-flow surface at $x_\text{min}$, the cylinder no-slip walls, and all remaining boundaries. 
At $\Omega_i$ we have $u_i = (u_\infty,0,0)'$ and we define the reference pressure as the average pressure on $\Omega_i$:
\begin{equation}
	p_\infty \ints{\Omega_i}{} = \ints{\Omega_i}{p}.
\end{equation}
With some manipulation of \eqref{eq:partial_KE_vol_int} we obtain the exact balance
\begin{align}
	\label{eq:partial_KE_vol_int_2}
   -\frac{1}{2} C_D A_\text{ref} u_\infty^3 = -P_D
  =
  \intv{ - \mathcal{P} - \mathcal{B}_\text{KE} - \mathcal{E} }  
  + \ints{
  \Omega_o}{ 
  	\underbrace{
    - \ol{u}_j n_j \tilde{E}_k 
    - \tilde{u}_i n_i (\ol{p} - p_\infty)
    + \frac{1}{\Rey_d}  \pdev{\tilde{E}_k}{x_j}n_j
    -  \tilde{u}_i  \ol{u_i'u_j'} n_j  
    }_{\widetilde{\mathcal{T}}_\text{KE}}
  },
\end{align}
where $P_D$ represents the power loss to drag, $\tilde{u}_i = \overline{u}_i - u_\infty \hat{x}_i$, and $\tilde{E}_k = \tfrac{1}{2} \tilde{u}_i \tilde{u}_i$.
The power lost to drag is dependent on both the drag coefficient, equal to the local drag coefficient \eqref{eq:local_drag} integrated over the cylinder surface, and the frontal reference area $A_\text{ref}$, equal to the cylinder diameter multiplied by the cylinder length.
The remaining transport terms are grouped by $\widetilde{\mathcal{T}}_\text{KE}$.
The mean kinetic energy balance can subsequently be written as
\begin{equation}
	\label{eq:KE_vol}
	0 =
      P_d
	- \langle \mathcal{P} \rangle_\mathcal{V}
	- \langle \mathcal{B}_\text{KE} \rangle_\mathcal{V}
	- \langle \mathcal{E} \rangle_\mathcal{V}
	+ \langle \widetilde{\mathcal{T}}_\text{KE} \rangle_{\Omega_o},
\end{equation}
where we have used the notation $\langle \cdot \rangle_\mathcal{V}$ to denote integration over volume $\mathcal{V}$, and $\langle \cdot \rangle_{\Omega_o}$ to denote integration over the surface $\Omega_o$.
From inspection of \eqref{eq:KE_vol} it is clear that direct integration of the KE transport equation \eqref{eq:KE_transport} over $\mathcal{V}$ leads to $\langle \mathcal{T}_\text{KE} \rangle_\mathcal{V} = \langle \widetilde{\mathcal{T}}_\text{KE} \rangle_{\Omega_o} + P_d$.

The remaining energy budgets, detailed in Appendix \ref{section:appendix_equations} equations \eqref{eq:PE_transport} to \eqref{eq:bb_transport}, do not require manipulation of the transport terms. Integrating the mean potential energy (PE) budget over $\mathcal{V}$ leads to
\begin{equation}
	\label{eq:PE_vol}
		 0 
		 =
		   \langle  \mathcal{B}_\text{KE}  \rangle_\mathcal{V}
		  +\langle  \mathcal{B}_\text{TKE} \rangle_\mathcal{V}
		  +\langle  \mathcal{T}_\text{PE} \rangle_\mathcal{V},
\end{equation}
where the mean potential energy $E_p = -\Ri_d \ol{\theta} z$ and $\mathcal{T}_\text{PE}$ represents PE transport.
The volume-integrated turbulent kinetic energy (TKE) budget is given as
\begin{equation}
	\label{eq:TKE_vol}
	0
	=
	 \langle \mathcal{P} 			\rangle_\mathcal{V}
	-\langle \mathcal{B}_\text{TKE} \rangle_\mathcal{V}
	-\langle \varepsilon 			\rangle_\mathcal{V}
	+\langle \mathcal{T}_\text{TKE} \rangle_\mathcal{V},
\end{equation}
where $\mathcal{T}_\text{TKE}$ represents the transport terms of TKE, $e_k = \ol{u_i' u_i'} / 2$.
While the budgets \eqref{eq:KE_vol} to \eqref{eq:TKE_vol} represents a closed system, irreversible mixing cannot be readily identified from the buoyancy fluxes $\mathcal{B}_\text{KE}$ and $\mathcal{B}_\text{TKE}$.
For this reason, we also assess the appropriately scaled transport equation for the bouyancy variance $\ol{b'b'}/2$, here termed the SBV budget (see Appendix \ref{section:appendix_equations} for details):
\begin{equation}
	\label{eq:TPE_vol}
	0
	=
	 \langle \mathcal{B}_\text{TKE} \rangle_\mathcal{V}
	+\langle \mathcal{B}_\text{TKE}^{xy} \rangle_\mathcal{V}
	-\langle \chi 			\rangle_\mathcal{V}
	+\langle \mathcal{T}_\text{SBV} \rangle_\mathcal{V}.
\end{equation}
Here, there is an additional buoyancy flux arising from horizontal mean buoyancy gradients, $\mathcal{B}_\text{TKE}^{xy}$, which become appreciable when in the `strongly stratified' wake regime, and $\mathcal{T}_\text{SBV} $ groups the various transport terms of the SBV equation \eqref{eq:bb_transport}.

\begin{figure}
	\centering
	\includegraphics{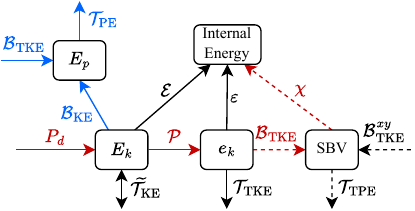}
	\caption{
		Volume integrated energy pathway for two-layer stably stratified flow past a vertical cylinder. 
		Terms correspond to volume-integrated mean kinetic energy ($E_k$), mean potential energy ($E_p$), turbulent kinetic energy ($e_k$) and scaled buoyancy variance (SBV) budgets: \eqref{eq:KE_vol}, \eqref{eq:PE_vol}, \eqref{eq:TKE_vol}, and \eqref{eq:TPE_vol} (note that the angled brackets have been removed for this schematic).
		The arrow directions have been inferred from the present simulation data.
		The critical pathway from power input due to drag ($P_d$) through to irreversible mixing of the buoyancy field ($\mathcal{X}$) is highlighted in red. 
		The KE sink due to stationary waves ($B_\text{KE}$) is highlighted in blue.
		The dashed lines highlight the uncertainty in interpretation of the SBV budget.
	}
	\label{fig:mixing_schematic}
\end{figure}

The four volume-integrated transport budgets are reported, schematically, in Figure \ref{fig:mixing_schematic}, which connects the four conserved energy reservoirs to the internal energy of the system, which increases through irreversible dissipative processes.
The arrow directions are determined from the computation of respective terms across all cases and volumes with $x_\mathcal{V} \geq 1$.
While many terms are not sign-definite, their signs appear robust in the volume integrals, with the exception of $\widetilde{\mathcal{T}}_\text{KE}$.
These grouped KE transport terms are very small compared to $P_d$ and are generally a sink of KE, although become a very slight KE source when stratification is strong.
This will be discussed further when analysing the buoyancy flux. 

It should also be noted that this system is not entirely closed due to the use of SBV to approximately identify irreversible mixing.
One particular issue is that $\mathcal{T}_\text{SBV}$ is not strictly a transport term, due to the spatial dependence of $N^2$.
If the domain were closed, or large enough to fully resolve mixing processes in the wake, the term would be minimal. 
Indeed, we find that when the integration volume is sufficiently large (large $x_\mathcal{V}$) the transport term is negligible.
An additional challenge is the extra source term $\mathcal{B}_\text{TKE}^{xy}$, which is not necessarily zero if there are appreciable horizontal gradients in $\ol{b}$. 
As a result, $\mathcal{X}$ is not necessarily smaller than $\mathcal{B}_\text{TKE}$, as it contains additional external sources. 
Finally, since the mean buoyancy flux is non-zero, it is important to consider PE transport, since this is an additional sink of KE, albeit a reversible one.
Regardless, this provides a framework with which the energy pathway from power loss to drag can be identified.

\begin{figure}
\centering
\includegraphics[width=\textwidth]{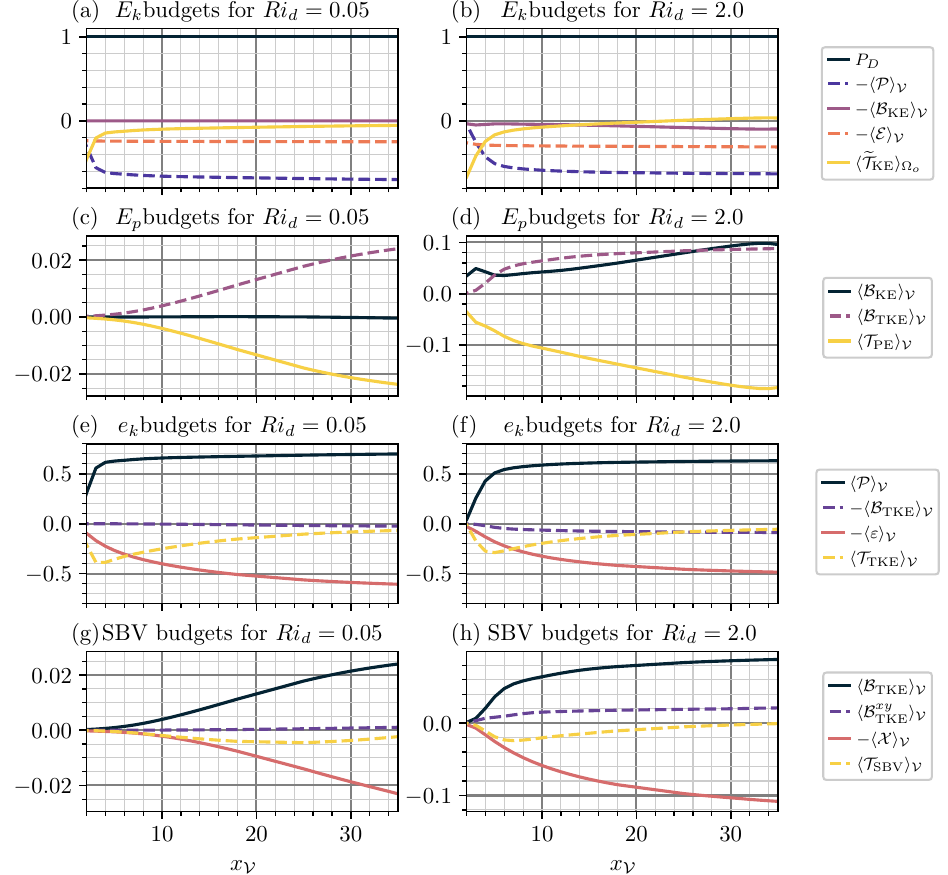}
\caption{
Volume integrated energy budgets for $\Ri_d = 0.05$ in panels (a,c,e,g) and $\Ri_d = 2.0$ in panels (b,d,f,h), all normalised by the power lost to drag, $P_d$.
$\Rey_d = 500$ for all data presented. 
Panels (a) and (b) show mean kinetic energy budgets \eqref{eq:KE_vol},
panels (c) and (d) show mean potential energy budgets \eqref{eq:PE_vol},
panels (e) and (f) show turbulent kinetic energy budgets \eqref{eq:TKE_vol},
and panels (g) and (h) show scaled buoyancy variance budgets \eqref{eq:TPE_vol}.
All budgets are presented as a function of $x_\mathcal{V}$, which specifies the maximum streamwise bounds of the integration volume $\mathcal{V}$.
}
\label{fig:int_energy_budgets}
\end{figure}

The four transport budgets are reported in Figure \ref{fig:int_energy_budgets} for \caseone\ and \casethree, as a function of $x_\mathcal{V}$.
Note that we have normalised all terms by the power input $P_d$.
The mean kinetic energy budget shows that $P_d$ is balanced primarily by TKE production and mean KE dissipation for \caseone, with a very small contribution from transport terms. 
The mean KE dissipation $\mathcal{E}$ is large due to the low Reynolds number, and peaks very close to the cylinder walls where mean flow gradients are largest. 
This is clear from Figure \ref{fig:int_energy_budgets} (a) since $\mathcal{E}$ is nearly independent of $x_\mathcal{V}$, indicating there are no mean dissipative processes in the far-wake. 
Nearly all TKE production occurs within the first few cylinder diameters, although there is some slow variation with $x_\mathcal{V}$ further downstream due to shear at the edges of the wake. 
With stronger stratification (\ref{fig:int_energy_budgets} (b)) we see that $\mathcal{P}$ is reduced, and $\mathcal{E}$ is increased, relative to the weaker stratification case (\ref{fig:int_energy_budgets} (a)). 
In addition, the mean buoyancy flux becomes important, and $\widetilde{\mathcal{T}}_\text{KE}$ becomes slightly positive.
The mean buoyancy flux reaches approximately 10\% of the power input from drag, although this is partially balanced by the net positive transport terms. 
There is a direct interplay between transport and mean buoyancy flux, due to the pressure contribution to $\widetilde{\mathcal{T}}_\text{KE}$ \eqref{eq:partial_KE_vol_int_2}.

The mean potential energy budgets (Figure \ref{fig:int_energy_budgets} (c) and (d)) unsurprisingly vary considerably in magnitude between \caseone\ and \casethree. 
We see that at low stratification $\mathcal{B}_\text{TKE}$ balances $\mathcal{T}_\text{PE}$ with negligible contributions from other terms.
The TKE buoyancy flux continues to increase throughout the full domain length, showing little sign of convergence, indicating that mixing processes are far from complete.
In contrast, \casethree\ shows both the mean and turbulent buoyancy fluxes contribute approximately equally to the PE budget.
The turbulent buoyancy flux rises quickly for $x_\mathcal{V} \lesssim 5$ before beginning to plateau as the wake advects further.
The mean buoyancy flux $\mathcal{B}_\text{KE}$ appears to grow as the wake propagates downstream, reaching its maximum value at $x_\mathcal{V} \approx 33$.
This is in part due to how the volume $\mathcal{V}$ is constructed, since the wave propagation direction is not aligned with the volume bounds. 
Once the stationary wave have developed, one could imagine the peaks and troughs effectively cancelling in the volume integral if the volume was constructed differently. 
If this were the case, a reduced $\mathcal{B}_\text{KE}$ would be compensated by an increase in KE transport terms ($\widetilde{\mathcal{T}}_\text{KE}$), and a reduction of mean potential energy; KE is either lost to an increase in PE, or leaves the integration volume through transport. 
Separation between KE transport and the mean buoyancy flux is therefore challenging.
Regardless, what is clear is that a significant proportion of the power source due to drag is captured by reversible fluxes and transport terms, and the fate of this energy beyond the numerical domain is unclear.

For low stratification we see TKE production is balanced by the TKE dissipation rate, with a small contribution from $\mathcal{T}_\text{TKE}$ due to turbulence exiting the volume $\mathcal{V}$ (Figure \ref{fig:int_energy_budgets} (e)).
Turbulent production appears to vary little for $x_\mathcal{V} \gtrsim 20$,  although this is due to the strong TKE production in the near-cylinder wake, which is much greater than the TKE production at the edges of the far wake.
A longer domain would be required to observe convergence of $\langle\varepsilon\rangle_\mathcal{V}$, due to the timescales over which produced TKE can dissipate. 
This is evidenced by the large $\mathcal{T}_\text{TKE}$ at small $x_\mathcal{V}$ which transport TKE downstream until it is dissipated.
At strong stratification (Figure \ref{fig:int_energy_budgets} (f)) we see qualitatively similar behaviour, although with a contribution from the turbulent buoyancy flux which increases with $x_\mathcal{V}$, reaching approximately 20\% of the TKE dissipation rate.
In addition, TKE production converges more slowly, due to the downstream shift of the KV.

The SBV buoyancy flux ($\mathcal{B}_\text{TKE}$) gradually increases with $x_\mathcal{V}$ at low stratification (Figure \ref{fig:int_energy_budgets} (g)).
This flux is primarily balanced by the irreversible destruction rate $\mathcal{X}$ with a small contribution from $\mathcal{T}_\text{SBV}$. 
At strong stratification (Figure \ref{fig:int_energy_budgets} (h)) the spanwise flux $\mathcal{B}_\text{TKE}^{xy}$ also becomes important, reaching approximately 25\% of $\mathcal{B}_\text{TKE}$.
This term arises due to the reasonably strong horizontal buoyancy gradients when the recirculation region is introduced (Figure \ref{fig:meanu_ynormal}).
While  $\mathcal{B}_\text{TKE}$ shows some sign of convergence, it does continue to increase throughout the numerical domain due to turbulent activity at the edges of the thermocline.
As a result, $\mathcal{X}$ does not converge within the numerical domain, reaching approximately 10\% of $P_d$ for $x_\mathcal{V} \gtrsim 30$. 
This partially contains contributions from $\mathcal{B}_\text{TKE}^{xy}$ and so cannot be fully associated with irreversible mixing of the potential energy field. 

\begin{figure}
\centering
\includegraphics[width=\textwidth]{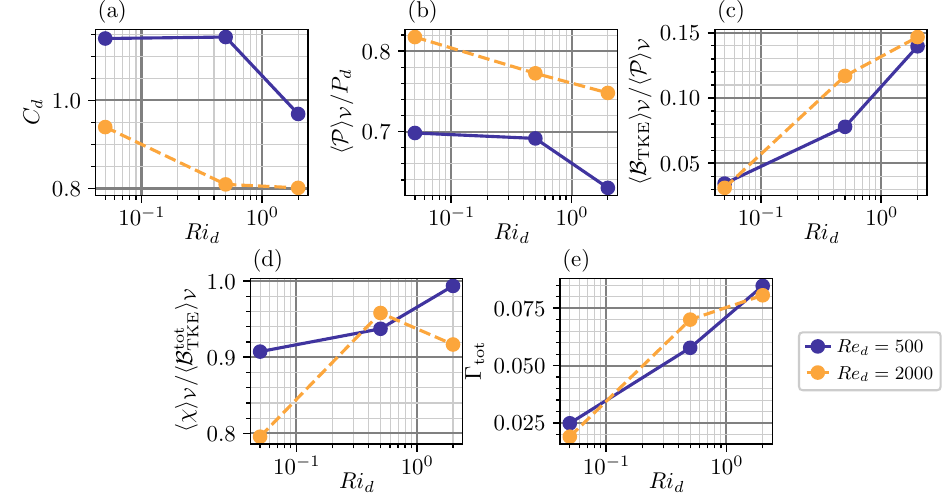}
\caption{
		The energy pathway ratios between power input from drag through to irreversible destruction of the scaled buoyancy variance, as a function of $\Ri_d$ and $\Rey_d$. 
		$\Gamma_\text{tot}$ is equal to the product of the four other ratios. 
		The volume-integrals are computed for $x_\mathcal{V} = 35$.
}
\label{fig:bulk_efficiencies}
\end{figure}

The energy pathway, from power input through to irreversible mixing of the buoyancy field, can be approximated by the product of four coefficients/efficiency ratios: $C_d$, $\langle \mathcal{P} \rangle_\mathcal{V} / P_d$, $\langle \mathcal{B}_\text{TKE}  \rangle_\mathcal{V} / \langle \mathcal{P} \rangle_\mathcal{V} $, and $\langle \mathcal{X} \rangle_\mathcal{V} / \langle \mathcal{B}_\text{TKE}^\text{tot} \rangle_\mathcal{V}$.
Here we have assumed that the ratio $\langle \mathcal{X} \rangle_\mathcal{V} / \langle \mathcal{B}_\text{TKE}^\text{tot} \rangle_\mathcal{V}$ is representative of the ratio between irreversible mixing of the buoyancy field, and the vertical turbulent buoyancy flux.
These ratios are reported for all six cases in Figure \ref{fig:bulk_efficiencies}, with budgets integrated over $\mathcal{V}$ with $x_\mathcal{V} = 35$.
Note that for the high Reynolds number cases we consider the imbalance in the SBV budget (a maximum of 5\%) an additional source for $\mathcal{X}$, due to sub-grid-scale dissipation.
The drag coefficient (Figure \ref{fig:bulk_efficiencies} (a)) is primarily a function of Reynolds number. 
Due to the moderate Reynolds numbers investigated, viscous forces differ considerably between $\Rey_d = 500$ and $\Rey_d = 2000$, accounting for the approximate 20\% difference in drag.
However, there is also a large reduction in drag as the flow transitions from the weakly stratified wake to the strongly stratified wake, arising due to the suppression of the KV and development of the large recirculation region. 

TKE production, as a proportion of the power input $P_d$ (Figure \ref{fig:bulk_efficiencies} (b)), increases with Reynolds number, by approximately 15\%, due to the reduced mean-flow dissipation rate.
There is also a Richardson number dependence that loosely corresponds to the regime change in the wake, where stronger stratification leads to a weaker TKE production, due both to the increased mean flow dissipation and the generation of the stationary internal waves. 
TKE production decreases from approximately 0.7 to 0.63 for $\Rey_d = 500$, as the flow undergoes the wake transition.

The flux Richardson number, given by the ratio between the buoyancy flux and TKE production (Figure \ref{fig:bulk_efficiencies} (c)),  appears insensitive to the Reynolds number, but varies strongly with $\Ri_d$.
However, it is unclear how the regime change affects this ratio, since there is a large difference in the buoyancy flux between \caseone\ and \casetwo, despite both being `weakly' stratified. 
Note that for all flows simulated here, the flux Richardson number is lower than the  accepted constant flux Richardson number $\Ri_f = 0.167$, particularly for the weak stratification cases.
It is unclear how this ratio may vary with a further increase in $\Ri_d$, but given the rate of change of these curves it would not be surprising to see $\Ri_f > 0.167$, even without considering local variations with depth. 
The ratio between the destruction rate of SBV and the total buoyancy flux (Figure \ref{fig:bulk_efficiencies} (d)) is greater than 90\% for all cases aside from \casefour, where vigorous mixing processes are far from complete as the wake advects out of the integration volume. 
Generally this ratio increases with increasing $\Ri_d$, aside from \casesix\ where it drops slightly. 
There is some degree of uncertainty regarding these higher Reynolds number cases given the approximated sub-grid-scale processes. 
However, these high ratios give confidence that these simulations capture the majority of mixing processes in the wake.

The total mixing efficiency, $\Gamma_\text{tot}$, equal to the product of the four ratios (Figures \ref{fig:bulk_efficiencies} (a) to (d)) is reported in Figure (Figure \ref{fig:bulk_efficiencies} (e).
$\Gamma_\text{tot}$ is a measure of how much energy from $P_d$ results in irreversible mixing of the buoyancy field.
The total mixing efficiency appears insensitive to Reynolds number, since the reduced drag coefficient at higher $\Rey_d$ is compensated by an increased TKE production ratio. 
However, it is strongly affected by $\Ri_d$,
varying between approximately 2\% at low $\Ri_d$ and increasing to approximately 8.5\% at high $\Ri_d$.

\section{Discussion}
\label{section:discussion}
This study has investigated the interactions between a uniform two-layer stratified flow and a vertically oriented cylinder, using DNS and high-resolution LES. 
The primary focus has been on understanding the influence of the flow Reynolds number $\Rey_d$ and Richardson number $\Ri_d$ on flow structure and mixing processes. 
We have found that a new regime emerges at sufficiently strong $\Ri_d$ where stationary internal waves develop at the interface between the unstratified and buoyancy-affected regions, as a result of strong vertical motions that are induced by recirculating flow structures within the thermocline.
These recirculating structures develop due to the vertical shearing across the buoyancy-interface, where strong buoyancy forces suppress momentum transfer in the lee of the cylinder and therefore increase the reattachment length.
The Karman Vortex is suppressed across the thermocline by these recirculating structures and the local drag coefficient is reduced. 
In contrast, the near-cylinder flow of the weakly-stratified regime is reasonably insensitive to the vertical coordinate, since mean shear is horizontal and therefore normal to gravity.

There are similarities between the standing waves of the strongly stratified wake and other natural flow phenomena. 
Mountain lee waves, generated by a stratified atmosphere passing topography, are visually similar, and are also generated by a stationary vertical disturbance in a stratified flow \citep{smith1979influence}.
Similarly, lee waves are also present in stratified flows past spheres due to the vertical disturbance of the structure, when stratification is sufficiently strong \citep{cocetta2021stratified}.
However, a key difference between the lee waves of mountains and spheres, and those of the vertically oriented cylinder herein, is that in the case of the vertical cylinders the perturbations are generated by the recirculation region that develops across the thermocline, rather than the (vertically uniform) physical structure itself.
In addition, the symmetrical nature of the vertical cylinder waves differs from typical lee waves, where the thermocline swells and contracts as the waves propagate, characteristic of `mode 2' solitary waves \citep{carr2015experiments}. 

The separation between the weakly and strongly stratified wake regimes is well characterised by the diagnostic parameters $\Ri_g$, $\Rey_b$, and $\Frou_h$:
The weakly stratified flow regime is characterised by $\Ri_g \gg 1$ everywhere, due to the negligible mean vertical shear.
Near the cylinder, we see $\Rey_b$ and $\Frou_h$ are also large, indicative of a weak influence from buoyancy on turbulent structures.
Vertical fluctuations, and therefore mixing of the buoyancy field, are initially small in the weakly stratified wake due to the dominance of horizontal motions; vertical fluctuations grow as the flow advects downstream and turbulence tends towards isotropy. 
At some point downstream, dependent on $\Rey_d$ and $\Ri_d$, turbulence decays and buoyancy forces become relevant, such that $\Rey_b$ and $\Frou_h$ become small.
The point where $\Frou_h\approx 1$ roughly corresponds to the peak in turbulent buoyancy flux, which occurs closer to the cylinder as $\Ri_d$ increases. 

In contrast, the strongly stratified wake is characterised by small $\Rey_b$ and $\Frou_h$ everywhere within the thermocline, aside from near the region of strong vertical flow, approximately four diameters downstream of the cylinder. 
In this region we also see $\Ri_g < 1$; vertical shearing in the flow becomes an important source of TKE within the thermocline and relatively efficient mixing processes develop.
Strong vertical fluctuations produced by vertical shearing are subsequently suppressed as the flow advects further downstream and turbulence is suppressed by buoyancy, increasing anisotropy. 
While the effect of the narrow and turbulent energetic wake on the temperature field is larger for weak stratification, we find that temperature perturbations are felt more widely for the strongly stratified wake, due to the stationary waves. 

\begin{figure}
\centering
\includegraphics{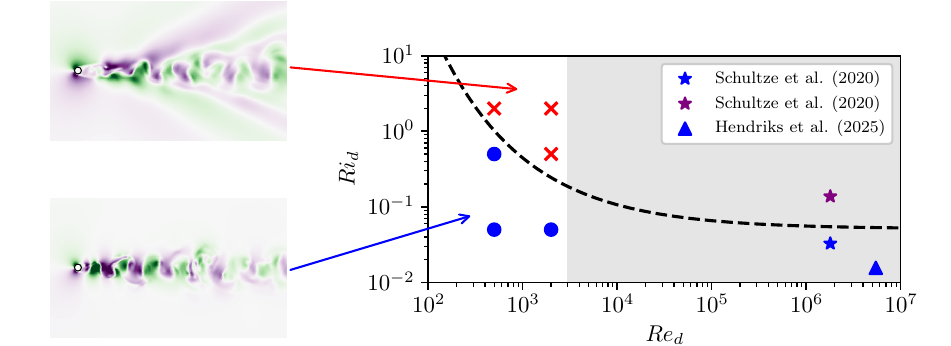}
\caption{
Regime diagram of two-layer stratified cylinder flow, with cases marked with a cross indicating cases with dominant stationary waves, and those marked with a circle indicating cases without.
The dashed line qualitatively represents the separation between these flow regimes, with the shaded region indicating high uncertainty of behaviour at high Reynolds numbers. 
Field data \citep{schultze2020increased,hendriks2025impact} have been added to constrain high Reynolds number behaviour. 
}
\label{fig:regime_diagram}
\end{figure}

The transition between the two flow regimes is Reynolds and Richardson number dependent. There is an earlier transition to the `strong' stratification regime when $\Rey_d$ is increased from 500 to 2000. 
We hypothesise that this is due to the stronger importance of three-dimensionality at higher Reynolds numbers, which are therefore more susceptible to vertical buoyancy forces. 
However, it is unclear how this process may scale to higher Reynolds numbers, which is critical for predicting mixing at field scale.
To understand this gap in scales we have constructed a regime diagram in Figure \ref{fig:regime_diagram}, where simulations resulting in a strongly stratified wake are marked with crosses, and weakly stratified wakes are marked with circles, as a function of $\Rey_d$ and $\Ri_d$.
The dashed line represents the approximate boundary between these two regimes, which we have extrapolated to small and large $\Rey_d$.
The low Reynolds number extrapolation is based upon the experimental results of \citet{meunier2012stratified}, who investigated the onset of instability in a uniformly stratified cylinder wake.
They found that the KV emerges independent of stratification strength for vertically aligned cylinders, at $\Rey_d \approx 45$, due to the lack of vertical motions that buoyancy may effect.
For this reason, we hypothesis that the boundary between two-layer weakly and strongly stratified wakes will occur at larger $\Ri_d$ as $\Rey_d$ decreases, until eventually the strongly stratified regime cannot develop, due to the lack of vertical motion.
The critical $\Ri_d$ value therefore decreases with $\Rey_d$, at least for low Reynolds numbers. 

At high Reynolds numbers there is much greater uncertainty, but the field data of \citet{schultze2020increased} and \citet{hendriks2025impact} can help constrain the problem.
Both these studies measured the wakes generated by tidal flows past monopiles in the German/Belgian regions of the North Sea. 
We have approximated $\Rey_d$ and $\Ri_d$ based on their maximum measured temperature differences, and the tidal velocities when data were recorded. 
Note however, that the field data differ from the idealised DNS, particularly due to background shear and turbulence.
There is a 3-order of magnitude difference between our simulation Reynolds numbers and those of the field. 
Despite this, we are confident that at least two of these field measurements lie in the `weakly stratified' wake regime, given \citet{schultze2020increased} and \citet{hendriks2025impact} report a highly energetic and localised turbulent wake extending far downstream of the structure.
However, the second field campaign of \citet{schultze2020increased} could not detect an energetic wake in their temperature measurements when the background stratification was stronger ($\Delta T \approx 2.1 \  {}^\text{o} \text{C}$, corresponding to $\Ri_d \approx 0.14$).
We hypothesise that this was due to the regime change between weak and strongly stratified wakes, where temperature variations would only be detectable closer to the structure, and large-scale stationary waves may be present. 
While \citet{schultze2020increased} did not report observing such large-scale structures, they may be more challenging to detect due to their spatial extent, steady-nature, and relatively weak strength.
If this hypothesis is correct, the Reynolds number dependence of the mixing regimes appears to plateau as $\Rey_d$ increases.
However, there is great uncertainty in this scaling. 
The LES carried out by \citet{schultze2020increased} did not demonstrate a clear regime change in their stratified wake simulations, although with only a 1 m vertical grid resolution it is unclear whether buoyancy length scales were adequately captured.
Indeed, \citet{schultze2020increased} note significant dissipation and mixing in their simulated background flows, even in the absence of infrastructure. 
It is therefore vital that future work addresses the scale gap between DNS and field observations, and extends analysis to flow profiles with vertical shear and background turbulence. 

It is also important to discuss how further increases in $\Ri_d$ may affect dynamics. 
If the vertical domain extent were large enough, the unstratified regions would be unaffected by any change in $\Ri_d$.
While high $\Ri_d$ may suppress turbulent activity across the thermocline, there must be some interface between the buoyancy affected region and the unstratified active turbulence. 
This will subsequently lead to the strong vertical shearing at the thermocline edges, and a suppression of the KV within the thermocline. 
For this reason we speculate that higher $\Ri_d$ will fall into the same `strongly stratified' regime identified herein, where vertical motions will develop in response to the vertical shearing, leading to waves propagating on the stratified-unstratified interfaces. 
Differences will occur regarding how deeply vertical flow structures may penetrate into the strongly stratified regions, which will subsequently limit the extent of vigorous mixing. 
However, we speculate the fundamental features that separate weakly and strongly stratified wakes will be maintained as $\Ri_d$ is increased.

Using the volume-integrated KE and PE budgets we have also identified the energy pathway between the power input associated with the cylinder drag, $P_d$, through to irreversible mixing of the buoyancy field, which we approximate using the destruction rate of the scaled buoyancy variance, $\mathcal{X}$.
The power input $P_d$ is proportional to the cylinder drag coefficient, which we find varies strongly with $\Rey_d$, due to the moderate Reynolds numbers, and $\Ri_d$, due to the suppression of turbulence across the thermocline when stratification is strong. 
The strongly stratified regime reduces drag by at least 15\% compared to the weakly stratified wake. 
The power loss from drag arises as a source term for time-averaged kinetic energy, which is balanced primarily by TKE production, although there are notable contributions from viscous KE dissipation (due to the moderate Reynolds numbers) and, in the case of the strongly stratified wake, the mean buoyancy flux and KE transport terms. 
The latter two arise due to the large-scale stationary waves that reversibly exchange energy between the KE and PE reservoirs.  
We show that this mean buoyancy flux can be as large as the turbulent buoyancy flux, although separating transport processes from this mean buoyancy flux is challenging and dependent on the construction of the integration volume.
The sum of the buoyancy flux and KE transport terms reaches approximately 8\% of $P_d$ when in the strongly stratified regime, doubling from approximately 4\% when stratification is weakest.
A key difference between these processes is that transport in the weakly stratified wake is concentrated in the narrow energetic wake which has not yet dissipated/mixed all the produced TKE from shear. In contrast, the strongly stratified wakes lead to wide-spread transport and buoyancy fluxes, due to stationary internal waves. 

The fraction of $P_d$ that is transferred to TKE is balanced primarily by the TKE dissipation rate, with a small contribution from the turbulent buoyancy flux. 
The ratio $\mathcal{B}_\text{TKE} / \mathcal{P}$ (the flux Richardson number) is strongly dependent on $\Ri_d$, varying from approximately 0.03 to 0.15, when the terms are integrated over the full domain volume. 
This is weaker than the typically assumed value of $\Ri_f \approx 0.167$ for mixing in stratified waters, although local values within the thermocline will be larger. 
The irreversible mixing rate is approximated by $\mathcal{X}$, which is at least 90\% of the turbulent buoyancy flux for five of the six cases investigated, generally increasing with increasing $\Ri_d$.
When $\Ri_d$ is weaker, transport is responsible for the imbalance, where mixing processes are not complete as the wake advects out of the domain. 
Indeed, the buoyancy flux and SBV destruction rate are still strongly dependent on the downstream extent of the integration volume, indicating that mixing processes are not constrained to the numerical domain. 

This work is motivated by the rapid expansion of offshore wind farms, which are now developing in deep seasonally stratified waters \citep{dorrell2022anthropogenic}. 
This is a recognised problem that is seeing increasingly active research  \citep{rennau2012effect,cazenave2016unstructured,christiansen2022emergence,christiansen2023large}, although all modelling strategies suitable for assessing regional scale impacts of shelf-sea developments are dependent on parameterising the energy pathway presented herein.
Current approaches have correctly identified the drag coefficient as a key uncertainty \citep{carpenter2016potential}; here we show that not only is it structure-dependent, but also strongly dependent on the stratified wake regime.
For accurate predictions of offshore wind impacts, it is crucial that this regime-dependence is incorporated into future parameterisations. 
In addition, we have shown that a considerable proportion of KE injected into the flow by vertical cylinders is lost to stationary internal waves. 
These waves could potentially propagate much longer distances than the energetic wake, eventually losing their energy through wave-breaking, or interactions with downstream structures.
This raises further questions regarding wake-wake interactions in large arrays of offshore wind turbines, since these waves could be expected to travel long distances.
Further work is required to extend high-fidelity simulations to resolve interactions between multiple wakes. 

Parameterisations have historically been calibrated against RANS data \citep{rennau2012effect}.
However, it is unclear whether the strongly stratified regime is able to be accurately captured by RANS/LES due to the high resolution required to capture over-turning scales.
Nevertheless, we note that the LES of \citet{schultze2020increased} obtained mixing efficiencies ($\gamma = \Ri_f / (1-\Ri_f)$) ranging between 0.09-0.14, which is of a similar range to the values obtained in this study (0.03-0.13).
Disagreements may be due to the different model set-up where background shear was incorporated into the model of \citet{schultze2020increased}, or could be due to the higher Reynolds numbers ($\Rey_d \approx 10^6$).
Regardless, both sets of simulations predict that the mixing efficiency is strongly dependent on the stratification strength.

There are a number of other uncertainties that require constraining in future work. 
In particular, mixing dynamics will be strongly affected by background shear and buoyancy profiles, background turbulence, wake-wake interactions, and of course the wider $\Rey_d$ and $\Ri_d$ regime space. 
In the field, further complications are introduced through atmospheric effects, other background turbulent processes, and seasonal/climate variability.
Of particular importance is the influence of structure design on mixing efficiency, as offshore wind infrastructure transition from monopiles to floating designs. 
In the $\Rey_d - \Ri_d$ parameter space herein, we show that the horizontal shearing associated with weakly-stratified cylinder wakes is not an efficient source of mixing until appreciable vertical fluctuations develop in the wake.
In contrast, more complex offshore wind infrastructure (e.g. floating semi-submersible structures) will directly create vertical shearing in the flow, in combination with horizontal motions. 
As a result, we anticipate that both the drag coefficient and flux-Richardson number will be larger, although still strongly dependent on stratification strength. 
Finally, we note that even in the case of vertical cylinders the drag coefficient and mixing dynamics vary strongly with depth.
It is not yet clear how such variations should be incorporated into oceanographic models. 
These considerations make it vital for future studies to be carried out that incorporate these additional processes, such that local and regional impacts of offshore renewable energy on stratified shelf seas can be accurately predicted.

\section{Conclusions}
This study has investigated the interactions between two-layer stably stratified flows and vertically oriented circular cylinders, using high-fidelity simulations, for the first time. 
Flow characteristics are found to be highly dependent on the flow Reynolds and Richardson numbers, with a new `strongly stratified' wake regime emerging at sufficiently high Richardson numbers.
The `weakly stratified' wake regime is characterised by a dominance of horizontal motion; vertical mixing of the buoyancy field develops further downstream of the most energetic part of the flow, once the flow has become less anisotropic. 
The strongly stratified wake regime is characterised by a large vertical recirculation region that develops across the thermocline, which leads to strong vertical shearing. 
This recirculation region leads to large-scale stationary internal waves which transport up to 10\% of the mean kinetic energy input away from the cylinder.
Vertical mixing develops in a fundamentally different way in the strongly stratified regime, where the vertical shearing leads to a highly efficient local source of mixing in the thermocline.
However, the strong buoyancy forces result in a suppression of vertical motions as the wake advects downstream, such that the overall impact on the temperature field is small.

This study is motivated by the development of offshore wind infrastructure in stratified shelf seas; the weakly stratified wake regime is characteristic of current relatively shallow water sites, while the strongly stratified regime will become increasingly common as offshore wind developments move further offshore. 
The emergence of the strongly-stratified wake regime explains why wakes are challenging to detect when stratification is strong \citep{schultze2020increased}.
In total, between 2\% and 8.5\% of the power input due to drag results in mixing of the buoyancy field, which increases with stratification strength. 
For the strongly stratified wakes, this is of a similar proportion to the energy transported by the stationary waves, the fate of which must be understood and parameterised in regional-scale models. 
Future work should therefore focus on constraining the Reynolds-Richardson number parameter space and incorporating additional background flow processes, in order to quantify field-scale mixing processes.

\section*{Acknowledgements}
CJL was supported by an Early Career Fellowship funded by the Leverhulme Trust.
RMD would like to acknowledge funding support from NERC Independent Research Fellow Grant NE/S014535/1.
We acknowledge the use of the Viper High Performance Computing facility of the University of Hull, and the Lovelace HPC service at Loughborough University.

\appendix
\titleformat{\section}
  {\normalfont\Large\bfseries}
  {Appendix \thesection:}{1em}{}

\numberwithin{equation}{section}

\section{Time-averaged transport equations}
\label{section:appendix_equations}
This Appendix details the time-averaged transport equations for mean kinetic energy (KE), mean potential energy (PE), turbulent kinetic energy (TKE), and the scaled buoyancy variance (SBV). 
Adopting summation notation, the transport equation for mean kinetic energy, ($E_k = \tfrac{1}{2} \ol{u}_i \ol{u}_i$), per unit volume, is obtained from the
dot-product of the velocity field and the momentum equation \eqref{eq:mom}:
\begin{equation}
\label{eq:KE_transport}
 0 =  
   \underbrace{ - \frac{1}{\Rey_d} \pdev{\ol{u}_i}{x_j} \pdev{\ol{u}_i}{x_j}}_{\substack{\text{KE dissipation} \\
    \vphantom{\text{Mean buoyancy flux}}  \\[4pt] \displaystyle -\mathcal{E}}}
  \underbrace{
	\vphantom{ \frac{1}{\Rey_d} \pdev{\ol{u}_i}{x_j} }
 	+ \ol{w} (\ol{b} - b_\text{in})
    }_{\substack{\text{KE buoyancy} \\ \text{flux} \\[4pt] \displaystyle -\mathcal{B}_\text{KE}}}
  \underbrace{+ \ol{u_i'u_j'} \pdev{\ol{u}_i}{x_j}}_{\substack{\text{TKE} \\ \text{production} \\[4pt] \displaystyle - \mathcal{P}}}
    \underbrace{
 	-  \pdev{}{x_j} \left( \ol{u}_j E_k  + \ol{u}_i \ol{p} \delta_{ij} - \frac{1}{\Rey_d}  \pdev{E_k}{x_j} +  \ol{u}_i \ol{u_i'u_j'}  \right)
    }_{\substack{\text{KE transport} \\
    \vphantom{\text{Mean buoyancy flux}}  \\[4pt] \displaystyle + \mathcal{T}_\text{KE}}},
\end{equation}
where the buoyancy $b = \Ri_d \theta$ with a reference inflow buoyancy $b_\text{in} = \Ri_d \theta_\text{in}$.
The first term, $\mathcal{E}$ represents viscous destruction of mean kinetic energy (KE), and is typically negligible for high Reynolds number flows \citep{Pope2000}. 
The second term, $\mathcal{B}_\text{KE}$ represents the mean buoyancy flux, which is neglected (or identically zero) in most flows concerning mixing of stratified flows where there is no mean vertical velocity.
The third term $\mathcal{P}$ represents production of turbulent kinetic energy (TKE), and the final term $\mathcal{T}_\text{KE}$ groups the various transport terms of the KE budget.

The KE transport equation is coupled to the transport of mean potential energy (PE), per unit volume, defined as $E_p = - \ol{b} z$.
PE transport is governed by
\begin{equation}
\label{eq:PE_transport}
     0 
     =
       \underbrace{
		\vphantom{\left( \frac{1}{\Rey_d} \pdev{\ol{u}_i}{x_j} \right) }
 	- \ol{w} (\ol{b} - b_\text{in})
    	}_{\substack{\text{KE buoyancy} \\ \text{flux} \\[4pt] \displaystyle + \mathcal{B}_\text{KE}}}
       \underbrace{
		\vphantom{\left( \frac{1}{\Rey_d} \pdev{\ol{u}_i}{x_j} \right) }
 	- \ol{b\rq{}w\rq{}}
    	}_{\substack{\text{TKE buoyancy} \\ \text{flux} \\[4pt] \displaystyle +\mathcal{B}_\text{TKE}}} 
       \underbrace{
     + \pdev{}{x_j} 
     	\left( 
     	-	\ol{u}_j E_p
     	- 	\ol{u}_j \Pi
       +    z \ol{b' u_j'} 
       + \frac{1}{\Pran \Rey_d} \pdev{E_p}{x_j}
       + \frac{2}{\Pran \Rey_d} b \hat{z}_j
     	\right)
     	}_{\substack{\text{PE transport} \\
    \vphantom{\text{Mean buoyancy flux}}  \\[4pt] \displaystyle + \mathcal{T}_\text{PE}}},
\end{equation}
where $\mathcal{B}_\text{TKE}$ represents the turbulent buoyancy flux and $\mathcal{T}_\text{PE}$ groups the PE transport terms.

KE and PE transport are coupled to the transport of turbulent kinetic energy (TKE), defined as $e_k = \tfrac{1}{2} \ol{u_i\rq{}u_i\rq{}}$:
\begin{equation}
\label{eq:TKE_transport}
0
=
  \underbrace{- \ol{u_i'u_j'} \pdev{\ol{u}_i}{x_j}}_{\substack{\text{TKE} \\ \text{production} \\[4pt] \displaystyle + \mathcal{P}}}
  \underbrace{
	\vphantom{\left( \frac{1}{\Rey_d} \pdev{\ol{u}_i}{x_j} \right) }
 	+ \ol{b\rq{}w\rq{}}
  }_{\substack{\text{TKE buoyancy} \\ \text{flux} \\[4pt] \displaystyle - \mathcal{B}_\text{TKE}}} 
  \underbrace{-  \frac{1}{Re}\ol{ \pdev{u'_i}{x_j} \pdev{u'_i}{x_j} }  }_{\substack{\text{TKE} \\ \text{dissipation} \\[4pt] \displaystyle -\varepsilon}}
   \underbrace{
		\vphantom{\left( \frac{1}{\Rey_d} \pdev{\ol{u}_i}{x_j} \right) }
   	 - \pdev{}{x_i} \left( \ol{u}_i e_k + \tfrac{1}{2}\ol{u_j' u_j' u_i'} + \ol{u_i' p'} - \frac{1}{Re}\pdev{e_k}{x_i} \right)
    }_{\substack{\text{TKE transport} \\
    \vphantom{\text{Mean buoyancy flux}}  \\[4pt] \displaystyle + \mathcal{T}_\text{TKE}}},
\end{equation}
where TKE transport terms are grouped into $\mathcal{T}_\text{TKE}$ and $\varepsilon$ represents the TKE dissipation rate. 
This system of transport equations 
\eqref{eq:KE_transport} to \eqref{eq:TKE_transport}
is a closed system, with the buoyancy flux and turbulent production terms transferring energy between the different budgets. 
However, a well recognised issue with these energy budgets is that mixing of the potential energy field cannot be discerned from the generally reversible fluxes $\mathcal{B}_\text{KE}$ and $\mathcal{B}_\text{TKE}$.
For this reason, we further assess budgets of the appropriately scaled buoyancy variance ($\tfrac{1}{2}\ol{b'b'}$) transport equation, the destruction rate of which is commonly adopted for diagnosing mixing of stratified flows \citep{taylor2019testing,smith2021turbulence,caulfield2021layering,zhou2024mixing}:
\begin{align}
	\label{eq:bb_transport}
\begin{split}
    0
    = 
     \underbrace{
	\vphantom{ \frac{1}{N^2 \Pran \Rey_d}\overline{\pdev{b'}{x_j}\pdev{b'}{x_j}} }
     	- \overline{w'b'}
     	}_{\substack{\text{TKE buoyancy} \\ \text{flux} \\[4pt] \displaystyle + \mathcal{B}_\text{TKE}}} 
     \underbrace{
	\vphantom{ \frac{1}{N^2 \Pran \Rey_d}\overline{\pdev{b'}{x_j}\pdev{b'}{x_j}} }
         - \frac{1}{N^2}\overline{u'b'}\pdev{\ol{b}}{x}    - \frac{1}{N^2}\overline{v'b'}\pdev{\ol{b}}{y}
	}_{\substack{\text{horizontal TKE} \\ \text{buoyancy flux} \\[4pt] \displaystyle + \mathcal{B}_\text{TKE}^{xy}}} 
    &  \underbrace{
    - \frac{1}{N^2 \Pran \Rey_d}\overline{\pdev{b'}{x_j}\pdev{b'}{x_j}}
    }_{\substack{\text{SBV destruction} \\ \text{rate} \\[4pt] \displaystyle - \mathcal{X}}}   \\ 
    &  \underbrace{
	\vphantom{ \frac{1}{N^2 \Pran \Rey_d}\overline{\pdev{b'}{x_j}\pdev{b'}{x_j}} }
    + \frac{1}{N^2} \pdev{}{x_j}\left[-\ol{u}_j \frac{1}{2}\ol{b\rq{}b\rq{}} + \frac{1}{\Pran \Rey_d}\pdev{}{x_j} \left(\frac{1}{2}\ol{b\rq{}b\rq{}}\right) - \frac{1}{2}\overline{b'b' u_j'}\right]
    }_{\substack{\text{SBV transport} \\
    \vphantom{\text{Mean buoyancy flux}}  \\[4pt] \displaystyle + \mathcal{T}_\text{SBV}}}
    ,
\end{split}
\end{align}
where $N^2 = \ol{b}_z$ is the buoyancy frequency.
The scaled buoyancy variance (SBV) transport equation is balanced by the TKE bouyancy flux $\mathcal{B}_\text{TKE}$, a spanwise turbulent buoyancy flux term $\mathcal{B}_\text{TKE}^{xy}$, the SBV (irreversible) destruction rate $\mathcal{X}$, and transport terms $\mathcal{T}_\text{SBV}$.
This form of the buoyancy variance transport is consistent with that of \citet{caulfield2021layering}, although several terms are present that can usually be neglected. 
In particular, the $x-y$ component of the turbulent buoyancy flux ($\mathcal{B}_\text{TKE}^{xy}$) is usually identically zero in stratified mixing problems (where $\ol{b}_z \gg \ol{b}_x, \ol{b}_y$) but is non-negligible in this flow due to inhomogeneity in the near cylinder wake, and subsequent emergence of relatively strong streamwise and spanwise buoyancy gradients. 
In a closed system, non-zero $\mathcal{B}_\text{TKE}^{xy}$ will result in $\mathcal{X} > \mathcal{B}_\text{TKE}$, which requires consideration when assessing mixing of the mean buoyancy field.
Nevertheless, $\mathcal{X}$ provides a local estimation of irreversible mixing. 

\bibliography{lit.bib}

@incollection{smith1979influence,
  title={The influence of mountains on the atmosphere},
  author={Smith, Ronald B},
  booktitle={Advances in geophysics},
  volume={21},
  pages={87--230},
  year={1979},
  publisher={Elsevier}
}

@article{cocetta2021stratified,
  title={Stratified flow past a sphere at moderate Reynolds numbers},
  author={Cocetta, Francesco and Gillard, Mike and Szmelter, Joanna and Smolarkiewicz, Piotr K},
  journal={Computers \& Fluids},
  volume={226},
  pages={104998},
  year={2021},
  publisher={Elsevier}
}

@article{bonnavion2022use,
  title={On the use of kinetic energy balance for the volumetric identification of drag sources of a blunt body. Application to road vehicles},
  author={Bonnavion, G and Bor{\'e}e, J and Herbert, V},
  journal={International Journal of Heat and Fluid Flow},
  volume={96},
  pages={108977},
  year={2022},
  publisher={Elsevier}
}

@article{gao2019energy,
  title={Energy-based drag breakdown in compressible flow by wake-plane integrals},
  author={Gao, An-Kang and Zou, Shufan and Shi, Yipeng and Wu, Jiezhi},
  journal={AIAA Journal},
  volume={57},
  number={8},
  pages={3231--3238},
  year={2019},
  publisher={American Institute of Aeronautics and Astronautics}
}

@article{brandt2014laboratory,
  title={Laboratory experiments on mass transport by large amplitude mode-2 internal solitary waves},
  author={Brandt, A and Shipley, KR},
  journal={Physics of Fluids},
  volume={26},
  number={4},
  year={2014},
  publisher={AIP Publishing}
}

@article{carr2015experiments,
    author = {Carr, M. and Davies, P. A. and Hoebers, R. P.},
    title = {Experiments on the structure and stability of mode-2 internal solitary-like waves propagating on an offset pycnocline},
    journal = {Physics of Fluids},
    volume = {27},
    number = {4},
    pages = {046602},
    year = {2015},
    month = {04},
    issn = {1070-6631},
}

@article{smith2021turbulence,
  title={Turbulence in forced stratified shear flows},
  author={Smith, Katherine M and Caulfield, CP and Taylor, JR},
  journal={Journal of Fluid Mechanics},
  volume={910},
  pages={A42},
  year={2021},
  publisher={Cambridge University Press}
}

@article{zhou2024mixing,
  title={Mixing in a strongly stratified turbulent wake quantified by bulk and conditional statistics},
  author={Zhou, Qi},
  journal={Journal of Fluid Mechanics},
  volume={997},
  pages={A41},
  year={2024},
  publisher={Cambridge University Press}
}

@article{taylor2019testing,
  title={Testing the assumptions underlying ocean mixing methodologies using direct numerical simulations},
  author={Taylor, JR and de Bruyn Kops, SM and Caulfield, CP and Linden, PF},
  journal={Journal of Physical Oceanography},
  volume={49},
  number={11},
  pages={2761--2779},
  year={2019}
}

@book{Pope2000,
  place={Cambridge},
  title={Turbulent Flows},
  publisher={Cambridge University Press},
  author={Pope, Stephen B.},
  year={2000}
}

@article{egbert2000significant,
  title={Significant dissipation of tidal energy in the deep ocean inferred from satellite altimeter data},
  author={Egbert, Gary D and Ray, Richard D},
  journal={Nature},
  volume={405},
  number={6788},
  pages={775--778},
  year={2000},
  publisher={Nature Publishing Group UK London}
}

@article{boyer1989linearly,
  title={Linearly stratified flow past a horizontal circular cylinder},
  author={Boyer, DL and Davies, PA and Fernando, HJS and Zhang, Xiuzhang},
  journal={Philosophical Transactions of the Royal Society of London. Series A, Mathematical and Physical Sciences},
  volume={328},
  number={1601},
  pages={501--528},
  year={1989},
  publisher={The Royal Society London}
}

@article{xu1995turbulent,
  title={Turbulent wakes of stratified flow past a cylinder},
  author={Xu, Yunxiu and Fernando, Harindra JS and Boyer, Don L},
  journal={Physics of Fluids},
  volume={7},
  number={9},
  pages={2243--2255},
  year={1995},
  publisher={American Institute of Physics}
}

@article{christin2021fluid,
  title={Fluid--structure interactions of a circular cylinder in a stratified fluid},
  author={Christin, Sarah and Meunier, Patrice and Le Diz{\`e}s, St{\'e}phane},
  journal={Journal of Fluid Mechanics},
  volume={915},
  pages={A97},
  year={2021},
  publisher={Cambridge University Press}
}

@article{bauer2013changing,
  title={The changing carbon cycle of the coastal ocean},
  author={Bauer, James E and Cai, Wei-Jun and Raymond, Peter A and Bianchi, Thomas S and Hopkinson, Charles S and Regnier, Pierre AG},
  journal={Nature},
  volume={504},
  number={7478},
  pages={61--70},
  year={2013},
  publisher={Nature Publishing Group UK London}
}

@article{wollast1998evaluation,
  title={Evaluation and comparison of the global carbon cycle in the coastal zone and in the open ocean. p.},
  author={Wollast, Roland},
  journal={The Sea, Vol. 10},
  pages={213--252},
  year={1998},
  publisher={Wiley}
}

@article{carpenter2016potential,
  title={Potential impacts of offshore wind farms on North Sea stratification},
  author={Carpenter, Jeffrey R and Merckelbach, Lucas and Callies, Ulrich and Clark, Suzanna and Gaslikova, Lidia and Baschek, Burkard},
  journal={PloS One},
  volume={11},
  number={8},
  pages={e0160830},
  year={2016},
  publisher={Public Library of Science San Francisco, CA USA}
}

@article{christiansen2022emergence,
  title={Emergence of large-scale hydrodynamic structures due to atmospheric offshore wind farm wakes},
  author={Christiansen, Nils and Daewel, Ute and Djath, Bughsin and Schrum, Corinna},
  journal={Frontiers in Marine Science},
  volume={9},
  pages={818501},
  year={2022},
  publisher={Frontiers Media SA}
}

@article{daewel2022offshore,
  title={Offshore wind farms are projected to impact primary production and bottom water deoxygenation in the North Sea},
  author={Daewel, Ute and Akhtar, Naveed and Christiansen, Nils and Schrum, Corinna},
  journal={Communications Earth \& Environment},
  volume={3},
  number={1},
  pages={292},
  year={2022},
  publisher={Nature Publishing Group UK London}
}

@article{rennau2012effect,
  title={On the effect of structure-induced resistance and mixing on inflows into the Baltic Sea: A numerical model study},
  author={Rennau, Hannes and Schimmels, Stefan and Burchard, Hans},
  journal={Coastal Engineering},
  volume={60},
  pages={53--68},
  year={2012},
  publisher={Elsevier}
}

@article{schultze2020increased,
  title={Increased mixing and turbulence in the wake of offshore wind farm foundations},
  author={Schultze, LKP and Merckelbach, LM and Horstmann, J and Raasch, S and Carpenter, JR},
  journal={Journal of Geophysical Research: Oceans},
  volume={125},
  number={8},
  pages={e2019JC015858},
  year={2020},
  publisher={Wiley Online Library}
}

@article{hendriks2025impact,
  title={The impact of offshore wind turbine foundations on local hydrodynamics and stratification in the Southern North Sea},
  author={Hendriks, Erik and Langedock, Kobus and Van Duren, Luca and Vanaverbeke, Jan and Boone, Wieter and Soetaert, Karline},
  journal={Frontiers in Marine Science},
  volume={12},
  pages={1619577},
  year={2025},
  publisher={Frontiers}
}

@article{cazenave2016unstructured,
  title={Unstructured grid modelling of offshore wind farm impacts on seasonally stratified shelf seas},
  author={Cazenave, Pierre William and Torres, Ricardo and Allen, J Icarus},
  journal={Progress in Oceanography},
  volume={145},
  pages={25--41},
  year={2016},
  publisher={Elsevier}
}

@article{floeter2017pelagic,
  title={Pelagic effects of offshore wind farm foundations in the stratified North Sea},
  author={Floeter, Jens and van Beusekom, Justus EE and Auch, Dominik and Callies, Ulrich and Carpenter, Jeffrey and Dudeck, Tim and Eberle, Sabine and Eckhardt, Andr{\'e} and Gloe, Dominik and H{\"a}nselmann, Kristin and others},
  journal={Progress in Oceanography},
  volume={156},
  pages={154--173},
  year={2017},
  publisher={Elsevier}
}

@article{austin2025enhanced,
  title={Enhanced bed shear stress and mixing in the tidal wake of an offshore wind turbine monopile},
  author={Austin, Martin J and Unsworth, Christopher A and Van Landeghem, Katrien JJ and Lincoln, Ben J},
  journal={Ocean Science},
  volume={21},
  number={1},
  pages={81--91},
  year={2025},
  publisher={Copernicus Publications G{\"o}ttingen, Germany}
}

@article{isaksson2025paradigm,
  title={A paradigm for understanding whole ecosystem effects of offshore wind farms in shelf seas},
  author={Isaksson, Natalie and Scott, Beth E and Hunt, Georgina L and Benninghaus, Ella and Declerck, Morgane and Gormley, Kate and Harris, Caitlin and Sj{\"o}strand, Sandra and Trifonova, Neda I and Waggitt, James J and others},
  journal={ICES Journal of Marine Science},
  volume={82},
  number={3},
  pages={fsad194},
  year={2025},
  publisher={Oxford University Press}
}

@article{zampollo2025does,
  title={Does the oceanographic response to wind farm wind-wakes affect the spring phytoplankton bloom?},
  author={Zampollo, Arianna and Murray, Rory O’Hara and Gallego, Alejandro and Scott, Beth},
  journal={Progress in Oceanography},
  pages={103512},
  year={2025},
  publisher={Elsevier}
}

@article{christiansen2023large,
  title={The large-scale impact of anthropogenic mixing by offshore wind turbine foundations in the shallow North Sea},
  author={Christiansen, Nils and Carpenter, Jeffrey R and Daewel, Ute and Suzuki, Nobuhiro and Schrum, Corinna},
  journal={Frontiers in Marine Science},
  volume={10},
  pages={1178330},
  year={2023},
  publisher={Frontiers Media SA}
}

@article{caulfield2021layering,
  title={Layering, instabilities, and mixing in turbulent stratified flows},
  author={Caulfield, CP},
  journal={Annual Review of Fluid Mechanics},
  volume={53},
  number={1},
  pages={113--145},
  year={2021},
  publisher={Annual Reviews}
}

@article{billant2000theoretical,
  title={Theoretical analysis of the zigzag instability of a vertical columnar vortex pair in a strongly stratified fluid},
  author={Billant, Paul and Chomaz, Jean-Marc},
  journal={Journal of Fluid Mechanics},
  volume={419},
  pages={29--63},
  year={2000},
  publisher={Cambridge University Press}
}

@article{lucas2019evolution,
  title={Evolution of oceanic near-surface stratification in response to an autumn storm},
  author={Lucas, Natasha S and Grant, Alan LM and Rippeth, Tom P and Polton, Jeff A and Palmer, Matthew R and Brannigan, Liam and Belcher, Stephen E},
  journal={Journal of Physical Oceanography},
  volume={49},
  number={11},
  pages={2961--2978},
  year={2019}
}

@article{bosco2014three,
  title={Three-dimensional instabilities of a stratified cylinder wake},
  author={Bosco, M and Meunier, P},
  journal={Journal of Fluid Mechanics},
  volume={759},
  pages={149--180},
  year={2014},
  publisher={Cambridge University Press}
}

@article{wang2024remote,
  title={Remote sensing unveils the explosive growth of global offshore wind turbines},
  author={Wang, Kechao and Xiao, Wu and He, Tingting and Zhang, Maoxin},
  journal={Renewable and Sustainable Energy Reviews},
  volume={191},
  pages={114186},
  year={2024},
  publisher={Elsevier}
}

@article{dorrell2022anthropogenic,
  title={Anthropogenic mixing in seasonally stratified shelf seas by offshore wind farm infrastructure},
  author={Dorrell, Robert M and Lloyd, Charlie J and Lincoln, Ben J and Rippeth, Tom P and Taylor, John R and Caulfield, Colm-cille P and Sharples, Jonathan and Polton, Jeff A and Scannell, Brian D and Greaves, Deborah M and others},
  journal={Frontiers in Marine Science},
  volume={9},
  pages={830927},
  year={2022},
  publisher={Frontiers Media SA}
}

@article{meunier2012stratified,
  title={Stratified wake of a tilted cylinder. Part 1. Suppression of a von K{\'a}rm{\'a}n vortex street},
  author={Meunier, Patrice},
  journal={Journal of Fluid Mechanics},
  volume={699},
  pages={174--197},
  year={2012},
  publisher={Cambridge University Press}
}

@article{williamson1996vortex,
   author = "Williamson, C H K",
   title = "Vortex Dynamics in the Cylinder Wake", 
   journal= "Annual Review of Fluid Mechanics",
   year = "1996",
   volume = "28",
   pages = "477-539",
   publisher = "Annual Reviews",
   issn = "1545-4479",
   type = "Journal Article",
}

@article{dong2014robust,
  title={A robust and accurate outflow boundary condition for incompressible flow simulations on severely-truncated unbounded domains},
  author={Dong, S and Karniadakis, GE and Chryssostomidis, C},
  journal={Journal of Computational Physics},
  volume={261},
  pages={83--105},
  year={2014},
  publisher={Elsevier}
}

@article{nek5000,
  title={Version 19.0},
  author={NEK5000},
  journal={Argonne National Laboratory, Illinois},
  publisher={Available: \url{https://nek5000.mcs.anl.gov}},
  year={2019}
}

@article{mittal2021multirate,
  title={Multirate timestepping for the incompressible Navier-Stokes equations in overlapping grids},
  author={Mittal, Ketan and Dutta, Som and Fischer, Paul},
  journal={Journal of Computational Physics},
  volume={437},
  pages={110335},
  year={2021},
  publisher={Elsevier}
}

@article{mittal2019nonconforming,
  title={Nonconforming Schwarz-spectral element methods for incompressible flow},
  author={Mittal, Ketan and Dutta, Som and Fischer, Paul},
  journal={Computers \& Fluids},
  volume={191},
  pages={104237},
  year={2019},
  publisher={Elsevier}
}

@article{lloyd2022coupled,
  title={The coupled dynamics of internal waves and hairpin vortices in stratified plane Poiseuille flow},
  author={Lloyd, CJ and Dorrell, RM and Caulfield, CP},
  journal={Journal of Fluid Mechanics},
  volume={934},
  pages={A10},
  year={2022},
  publisher={Cambridge University Press}
}

@article{thorpe2016layers,
  title={Layers and internal waves in uniformly stratified fluids stirred by vertical grids},
  author={Thorpe, SA},
  journal={Journal of Fluid Mechanics},
  volume={793},
  pages={380--413},
  year={2016},
  publisher={Cambridge University Press}
}

@article{smyth2000length,
  title={Length scales of turbulence in stably stratified mixing layers},
  author={Smyth, William D and Moum, James N},
  journal={Physics of Fluids},
  volume={12},
  number={6},
  pages={1327--1342},
  year={2000},
  publisher={American Institute of Physics}
}

@incollection{orszag1979spectral,
  title={Spectral methods for problems in complex geometrics},
  author={Orszag, Steven A},
  booktitle={Numerical methods for partial differential equations},
  pages={273--305},
  year={1979},
  publisher={Elsevier}
}

@misc{canuto2012spectral,
	title={Spectral Methods in Fluid Dynamics},
	author={Canuto, C. and Hussaini, M.Y. and Quarteroni, A. and Thomas, A. JR.},
	year={2012},
	publisher={Springer}
}

@article{tomboulides1997numerical,
  title={Numerical simulation of low Mach number reactive flows},
  author={Tomboulides, AG and Lee, JCY and Orszag, SA},
  journal={Journal of Scientific Computing},
  volume={12},
  pages={139--167},
  year={1997},
  publisher={Springer}
}

\end{document}


\onehalfspacing

\maketitle
\noindent
This document contains supplementary material for the article:
\begin{enumerate}
	\item[] Lloyd, C.J. and Dorrell, R.M., 2025. Mixing by offshore wind infrastructure: Resolving the density stratified wakes past vertical cylinders. arXiv preprint arXiv:2512.10751.
\end{enumerate}
Some additional figures have been included in this supplementary material to provide further information on the overlapping SSEM grid construction, show some additional data and visualisations, and provide further comparisons between low and high Reynolds number simulations.

\begin{figure}
	\centering
	\def\scaley{0.5}
	\begin{tikzpicture}[x=0.5\textwidth,y=0.5\textwidth*\scaley,scale=0.8]
	    \node (pic1) at (0.5,-0.8) {\includegraphics[width=0.3\textwidth]{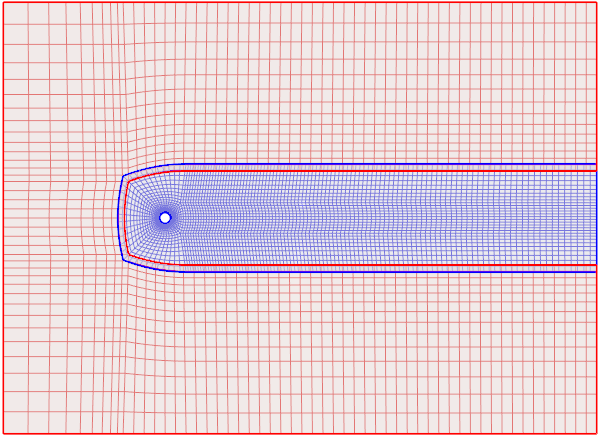}};
	    \node (pic2) at (1.5,-0.8) {\includegraphics[width=0.3\textwidth]{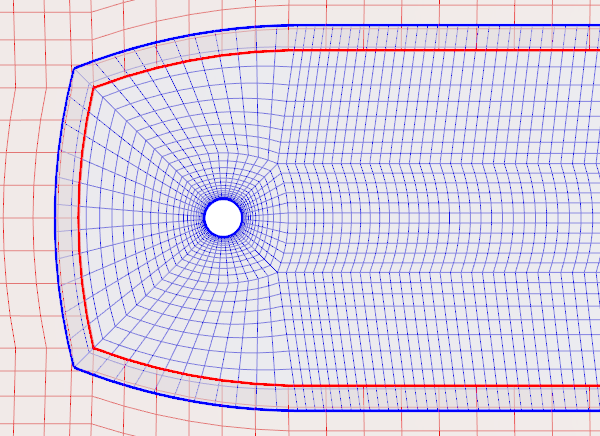}};
	    \node at (0.05,-0.2) {(a)};
	    \node at (1.05,-0.2) {(b)};
	\end{tikzpicture}
	\caption{
		Plan view of the overlapping spectral element grids (a), and a close-up view of mesh-refinement near the cylinder (b).
		The spectral element grids overlap by approximately one element. 
	}
\end{figure}

\begin{figure}
	\centering
	\includegraphics[width=\textwidth]{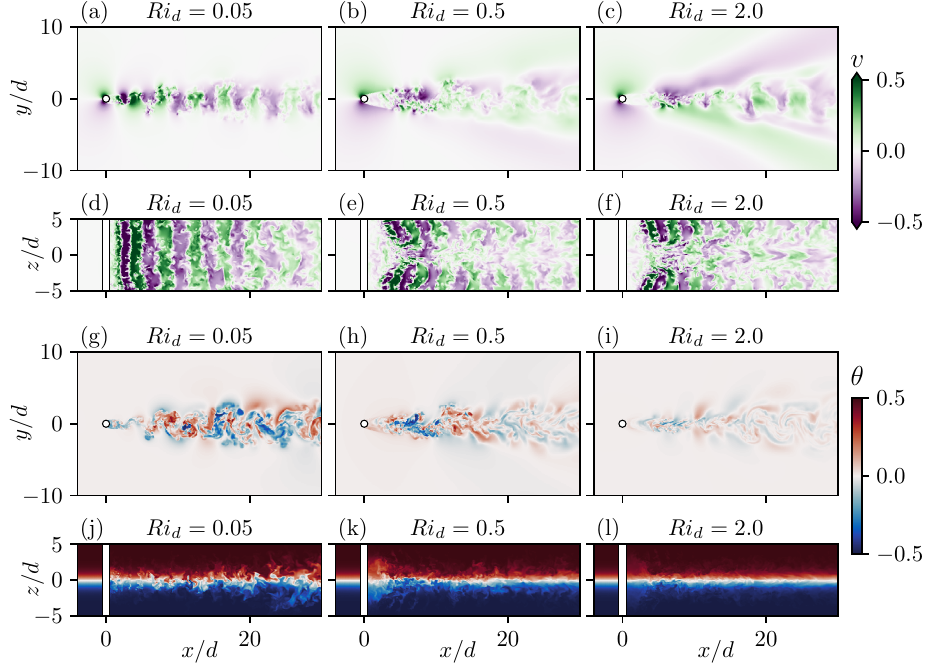}
	\caption{
		Visualisation of instantaneous spanwise velocity (a to f) and temperature (g to l). $\Rey_d = 2000$ for all panels with: $\Ri_d = 0.05$ in panels (a, d, g, j); $\Ri_d = 0.5$ in panels (b, e, h, k); and $\Ri_d = 2.0$ in panels (c, f, i, l).
		Panels (a) to (c) and (g) to (i) show a $z-$normal slice at $z=0$, and panels (d) to (f) and (j) to (l) show a $y-$normal slice at $y=0$.
		This figure is provided as a comparison against the low Reynolds number visualisations of Fig. 2 in Lloyd and Dorrell (2026).
	}
\end{figure}

\begin{figure}
	\centering
	\includegraphics[width=\textwidth]{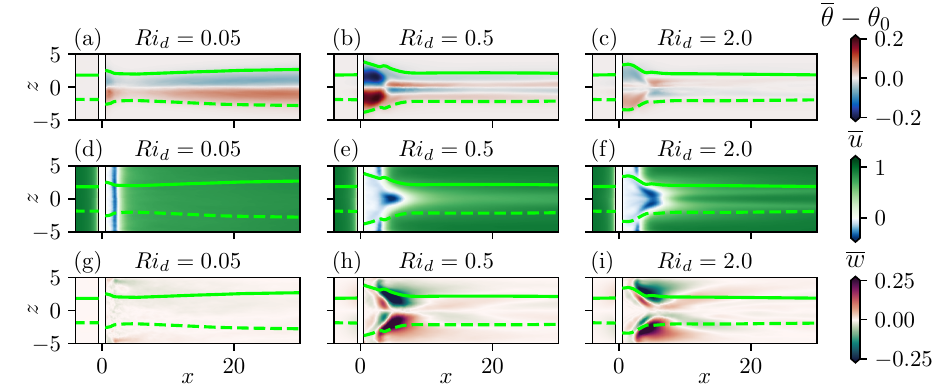}
	\caption{
		Time-averaged temperature (a to c), streamwise velocity (d to f) and vertical velocity (g to i) on a $y-$normal slice at $y=0$.
 		$\Rey_d = 2000$ for all panels with: $\Ri_d = 0.05$ in panels (a, d, g); $\Ri_d = 0.5$ in panels (b, e, h); and $\Ri_d = 2.0$ in panels (c, f, i).
		Temperature data are presented as a perturbation from the spatially varying background field $\theta_0(x,z)$.
		Lines represent the thermocline bounds, quantified by the contours $\overline{\theta} = \pm 0.475$.
		This figure is provided as a comparison against the low Reynolds number visualisations of Fig. 4 in Lloyd and Dorrell (2026).		
	}
\end{figure}

\begin{figure}
	\centering
	\includegraphics[width=\textwidth]{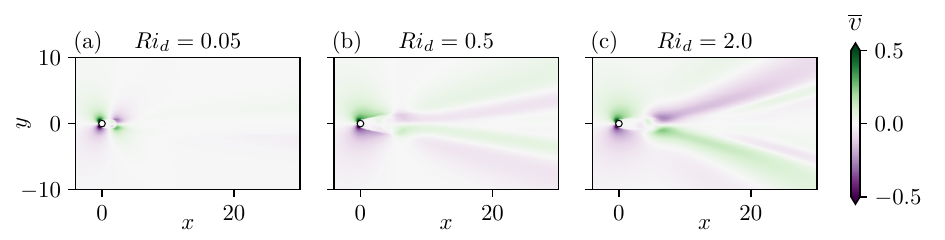}
	\caption{
		Time-averaged spanwise velocity on a $z-$normal slice at $z=0$.
 		$\Rey_d = 500$ for all panels with: $\Ri_d = 0.05$ in panel (a); $\Ri_d = 0.5$ in panel (b); and $\Ri_d = 2.0$ in panel (c).
		This figure is provided as a comparison against the low Reynolds number visualisations of Fig. 5 in Lloyd and Dorrell (2026).		
	}
\end{figure}

\begin{figure}
	\centering
	\includegraphics[width=\textwidth]{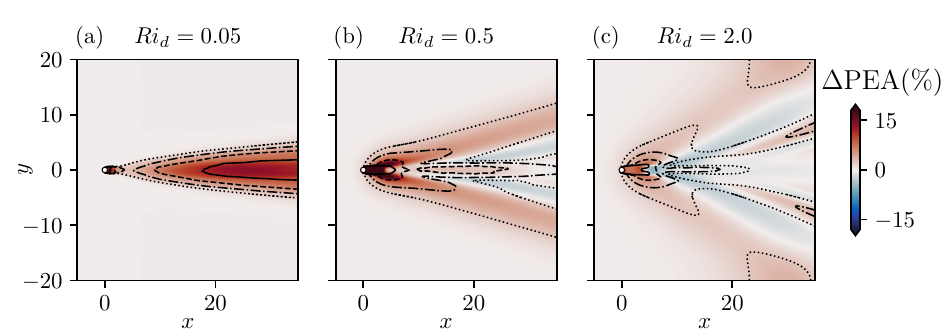}
	\caption{
		Change in the Potential Energy Anomaly, defined as $\int_{-2}^2 (E_p - E_{p0}) \  \mathrm{d} z / \left|\int_{-2}^2 E_{p0} \  \mathrm{d} z \right|$.
		$E_{p0}$ represents the spatially varying PE based upon the background temperature field $\theta_0(x,z)$.
		Contours represent the relative changes in thermocline width, defined as the vertical distance between the contours $\overline{\theta} = \pm 0.475$, 
		relative to respective spatially varying background values. 
		Linestyles represent a thermocline width change of 5\% (dash-dotted), 10\% (dashed), and 20\% (solid).
		$\Rey_d = 500$ for all panels with: $\Ri_d = 0.05$ (a); $\Ri_d = 0.5$ (b); and $\Ri_d = 2.0$ (c).
		This figure is provided as a comparison against the low Reynolds number visualisations of Fig. 8 in Lloyd and Dorrell (2026).
	}
\end{figure}

\begin{figure}
	\centering
	\includegraphics[width=\textwidth]{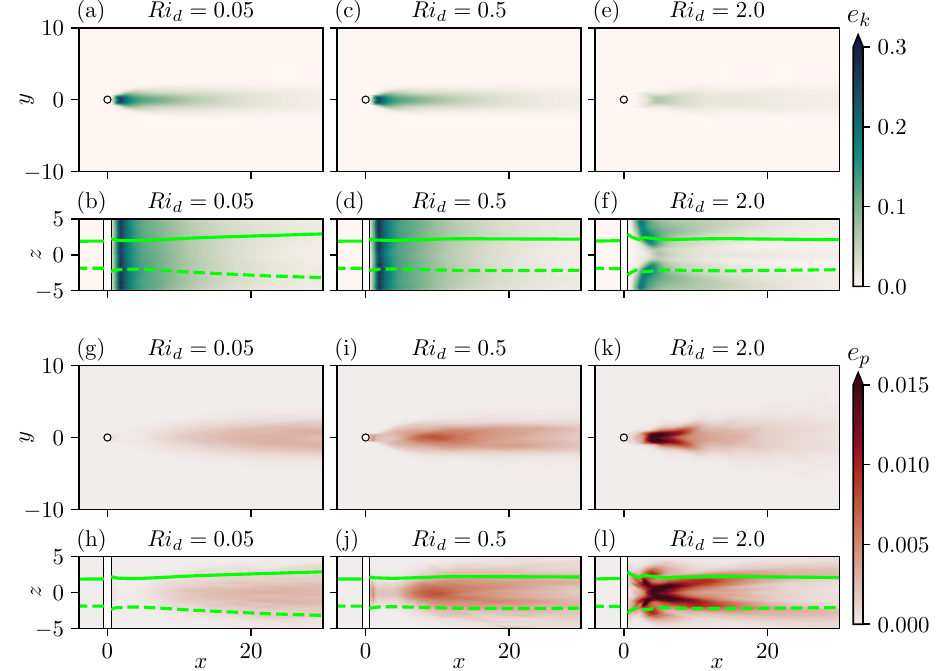}
	\caption{
		Visualisation of turbulent kinetic energy and turbulent potential energy on a $z-$normal slice at $z=0$ (panels a to c) and a $y-$normal slice at $y=0$ (panels d to f).
 		$\Rey_d = 500$ for all panels with: $\Ri_d = 0.05$ in panels (a) and (d); $\Ri_d = 0.5$ in panels (b) and (e); and $\Ri_d = 2.0$ in panel (c) and (f).
	}
\end{figure}

\begin{figure}
	\centering
	\includegraphics[width=\textwidth]{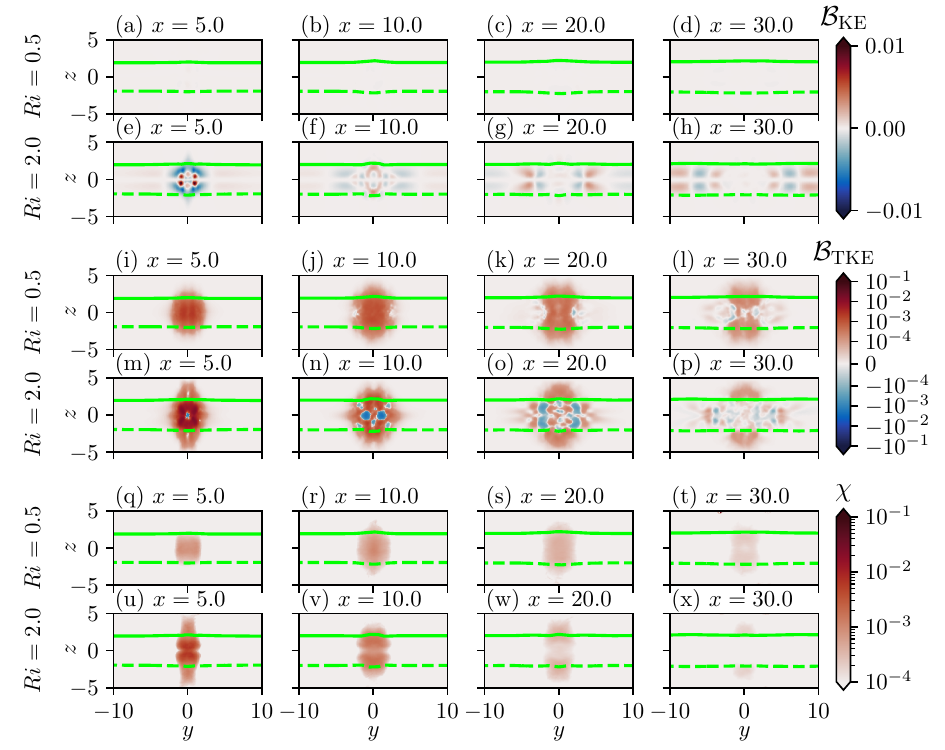}
	\caption{
		Visualisation of mean buoyancy flux (panels (a to h)), turbulent buoyancy flux (panels (i to p)), and turbulent potential energy destruction (panels (q to x)), on four $x-$normal slices, for $\Ri_d = 0.5$ and $\Ri_d = 2.0$.
		$\Rey_d = 500$ for all panels. 
		panels (a,e,i,m,q,u) visualise a slice at $x=5$,
		panels (b,f,j,n,r,v) visualise a slice at $x=10$,
		panels (c,g,k,o,s,w) visualise a slice at $x=20$,
		and panels (d,h,l,p,t,x) visualise a slice at $x=30$.
		Data for $\Ri_d = 0.5$ are presented in panels (a,b,c,d,i,j,k,l,q,r,s,t), 
		and data for $\Ri_d = 2.0$ are presented in panels (e,f,g,h,m,n,o,p,u,v,w,x). 
	}
\end{figure}

\begin{figure}
	\centering
	\includegraphics[width=\textwidth]{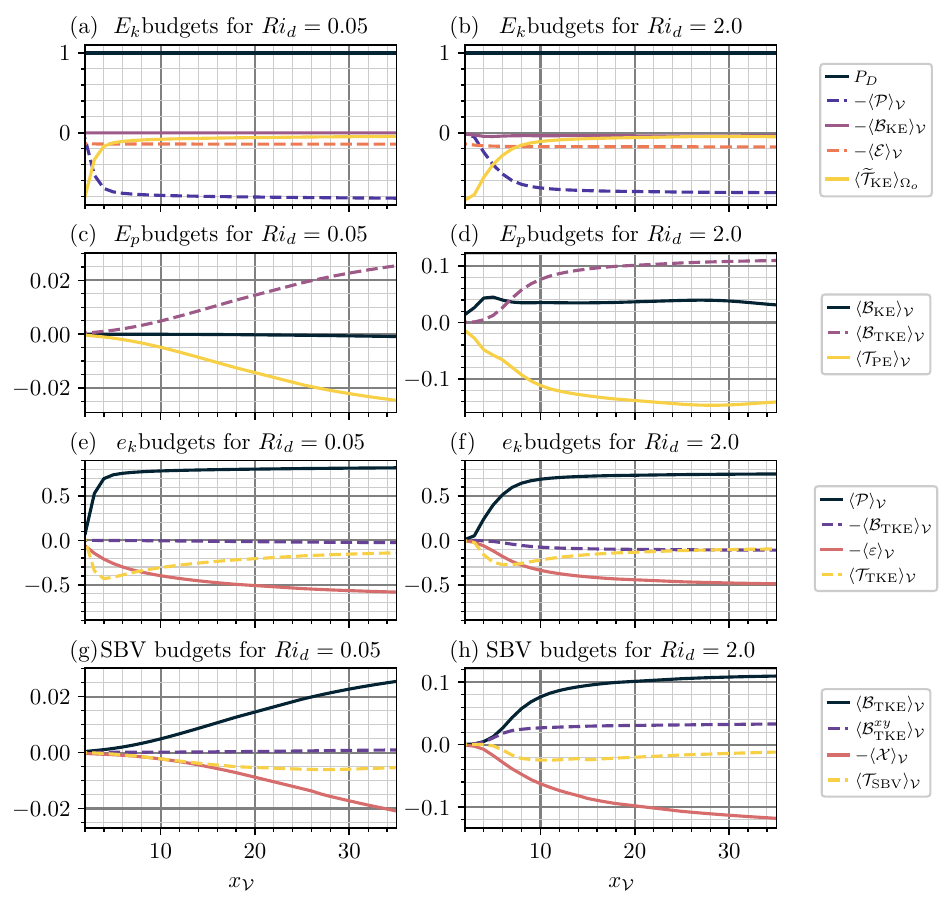}
	\caption{
		Volume integrated energy budgets for $\Ri_d = 0.05$ in panels (a,c,e,g) and $\Ri_d = 2.0$ in panels (b,d,f,h), all normalised by the power lost to drag, $P_d$.
		$\Rey_d = 2000$ for all data presented. 
		Panels (a) and (b) show mean kinetic energy budgets,
		panels (c) and (d) show mean potential energy budgets,
		panels (e) and (f) show turbulent kinetic energy budgets,
		and panels (g) and (h) show scaled buoyancy variance budgets.
		All budgets are presented as a function of $x_\mathcal{V}$, which specifies the maximum streamwise bounds of the integration volume $\mathcal{V}$.
		This figure is provided as a comparison against the low Reynolds number visualisations of Fig. 18 in Lloyd and Dorrell (2026).
	}
\end{figure}